\documentclass[11pt]{article}

\usepackage[final]{acl}

\usepackage{times}
\usepackage{latexsym}

\usepackage[T1]{fontenc}

\usepackage[utf8]{inputenc}

\usepackage{microtype}

\usepackage{inconsolata}

\usepackage{graphicx}

\usepackage{amsmath,amssymb,amsfonts}
\usepackage{textcomp}
\usepackage{xcolor}
\usepackage{booktabs}
\usepackage{multirow}
\usepackage{algorithm}
\usepackage{algorithmic}
\usepackage{hyperref}
\usepackage{bm}
\usepackage{url}
\usepackage{pifont}
\usepackage[most]{tcolorbox}
\usepackage{marvosym}
\title{EmoTra-TTS: Smooth Intra-Utterance Emotion Transitions \\ for Speech Synthesis}

\author{
 \textbf{Tianchi Liu\textsuperscript{1,$\dagger$}},
 \textbf{Zeyang Song\textsuperscript{2,$\dagger$}},
 \textbf{Tianrui Wang\textsuperscript{3}},
 \textbf{Zhipeng Li\textsuperscript{1}},
\\
 \textbf{Chenglin Xu\textsuperscript{1}},
 \textbf{Yiwen Guo\textsuperscript{4}}
\\
\\
 \textsuperscript{1}LIGHTSPEED,
 \textsuperscript{2}National University of Singapore, \\
 \textsuperscript{3}Nanyang Technological University,
 \textsuperscript{4}Independent Researcher
\\
 \small{
   $^\dagger$\textbf{Corresponding Authors:} \href{mailto:email@domain}{\{tianchi\_liu, zeyang\_song\}@u.nus.edu}
}
}

\begin{document}
\maketitle
\begin{abstract}
Psychological research on emotion dynamics has established that human affect is a continuous, evolving process: emotions rise, decay, and transition within seconds.
Current emotional text-to-speech (TTS) systems, however, condition on a single discrete label or static embedding per utterance, fundamentally misaligning with the temporal nature of affect.
While recent LLM-based TTS systems may implicitly vary prosody through text understanding, such variation is neither explicitly controllable nor precise enough for targeted intra-utterance transitions.
We address three challenges:
(1)~a \emph{multi-pass flow blending} pipeline synthesizes frame-aligned transition audio, circumventing the scarcity of natural intra-utterance transitions;
(2)~\emph{dual-stage Valence–Arousal–Dominance (VAD) conditioning} guides prosodic planning in the LLM and acoustic realization in the flow decoder via frame-level VAD embeddings;
(3)~\emph{direction--magnitude decoupled injection} structurally separates emotion direction from injection magnitude, preventing content degradation.
\textbf{EmoTra-TTS}\footnote{Demo page: \url{https://liu-tianchi.github.io/EmoTra_DemoPage/}; The complete code, including data synthesis and training pipelines: \url{https://github.com/Liu-Tianchi/EmoTra-TTS}} adds only \textbf{$+0.43\%$} parameters with no latency overhead, achieves \textbf{30\%–87\%} relative improvement on emotion transition quality, corroborated by \textbf{64.4\%–79.5\%} overall win rates in pairwise preference tests against four SOTA baselines and two commercial systems.

\end{abstract}

\section{Introduction}
\label{sec:intro}

Human emotion is not a static label but a \emph{dynamic process} that unfolds over time.
Research on \emph{emotion dynamics}~\cite{davidson1998,kuppens2017emotion} has established that affective states fluctuate continuously, exhibiting inertia, gradual decay, and smooth transitions between categories~\cite{verduyn2009, cowen2017self}, with temporal envelopes that carry communicative intent~\cite{trampe2015}.
Russell's circumplex model~\cite{russell1980} and the PAD framework~\cite{mehrabian1996} formalize this continuity by representing emotions in a continuous VAD space, supported by neuroscientific evidence~\cite{posner2005} and continuous gradients between categories~\cite{cowen2017self}.

These findings motivate emotion control that is continuous in both emotion space and time, especially for expressive speech scenarios such as storytelling, dubbing, and audiobook narration, where abrupt transitions are highly perceptible.

Despite this foundation, emotional TTS predominantly assigns one categorical label per utterance~\cite{lei2022msemotts}.
Recent systems offer finer control: EmoSphere-TTS~\cite{emosphere} uses spherical VAD vectors, EmoKnob~\cite{emoknob} provides disentangled style knobs, and WeSCon~\cite{wescon} explores discrete word-level control.
Yet these utterance- and word-level systems do not model the continuous temporal dynamics of emotion: the rise, transition, and decay that define natural affective experience.
Recent LLM-based TTS systems~\cite{Qwen3TTS, cosyvoice3, koeltts, prompttts2, indextts2, li2026restyle, wang2026evaluating, song2026diagnose} may implicitly vary prosody through text understanding, but such variation is neither user-controllable nor sufficiently precise for targeted emotion transitions within an utterance.

Bridging this gap requires addressing three interconnected challenges:
\textbf{(1) Data scarcity}: 
naturally occurring intra-utterance emotion transitions are severely underrepresented in emotional speech corpora, and eliciting them from actors is notoriously difficult~\cite{busso2008iemocap,zhou2022emotional, michel2026expressive}.
\textbf{(2) Dual-level conditioning}: modern LLM-based TTS~\cite{cosyvoice2,valle,seedtts} decomposes synthesis into prosodic planning (LLM) and acoustic realization (flow decoder), each requiring temporally aligned emotion signals~\cite{scherer2003}.
\textbf{(3) Magnitude control}: injecting conditioning into a pretrained decoder risks pushing it outside its operating range; learnable constraints are circumvented by co-adaptation~\cite{locatello2019challenging}.
To address these challenges, we propose \textbf{EmoTra-TTS} built upon three key components:
\begin{itemize}
    \item \textbf{Synthetic emotion transition data} (\S\ref{sec:data}): a multi-pass flow blending pipeline generates frame-aligned transition audio, circumventing the difficulty of recording natural transitions.

    \item \textbf{Dual-stage VAD conditioning} (\S\ref{sec:llm}, \S\ref{sec:flow}): VAD tokens condition the LLM for prosodic planning; frame-level VAD embeddings modulate the flow decoder's speaker pathway.

    \item \textbf{Direction--magnitude decoupled injection} (\S\ref{sec:method_injection}): \emph{LayerNorm + fixed scale} decouples emotion direction (learned) from injection magnitude (kept bounded via re-normalization), mitigating the content-expressiveness trade-off.
\end{itemize}

\section{Related Work}
\label{sec:related}

\subsection{Emotion Dynamics and Emotional TTS}

Emotion dynamics research~\cite{davidson1998,kuppens2017emotion} has established that affect possesses an intrinsic temporal microstructure.
Dimensional models~\cite{mehrabian1996} with acoustic correlates~\cite{scherer2003,banse1996} motivate VAD {transitions}, not merely VAD {points}, as conditioning signals.

Emotional TTS has evolved from discrete emotion embeddings~\cite{wang2018style} and prosody transfer~\cite{skerry2018towards} to continuous VAD control~\cite{emosphere,11049047,zhou2026emotional}.
EmoShift~\cite{emoshift} and CoCoEmo~\cite{cocoemo} explore activation steering, while EMORL-TTS~\cite{EMORLTTS} applies reinforcement learning for emotion control.
Recent efforts further explore finer-grained control: \citet{10832181} condition a flow-matching decoder on frame-level arousal--valence from reference audio, \citet{liang2026segment} explore segment-level modeling for intra-utterance emotion variation, and WeSCon~\cite{wescon} introduces word-level discrete control.
While these works address related forms of fine-grained control, explicit and controllable intra-utterance transitions along continuous VAD dimensions, specified as a user-defined trajectory rather than reference-driven or discrete control, remain underexplored~\cite{cui2025speechlm,controllable_survey}.
Recent benchmarks report the same limitation for commercial systems: InstructTTSEval~\cite{huang2025instructttseval} shows Gemini and VoxInstruct handle only rudimentary changes (e.g., voice rise), not genuine emotion change; EmergentTTS-Eval~\cite{manku2026emergentttseval} lists natural shifts to contrasting emotions among its most challenging cases; EmoS~\cite{wang2026emos} finds commercial models (e.g., GPT-4o-Audio, Gemini 2.5 Pro) lag far behind humans on cross-turn emotion transitions; and MINT-Bench~\cite{chen2026mint} identifies paralinguistic control, including explicit intra-utterance emotion transitions, as a major bottleneck for commercial systems (e.g., Gemini, GPT-4o mini TTS, ElevenLabs).
In contrast, EmoTra-TTS models such continuous transitions.

\subsection{Synthetic Data for Speech Tasks}
Synthetic data has proven effective for speech tasks: TTS-augmented ASR training~\cite{10889894}, speech translation~\cite{zhao-etal-2023-generating}, speech security~\cite{hwang2026scores, 10832142, 11222612}, speaker recognition~\cite{11249009, 11249126}, and emotion recognition~\cite{ma2024leveraging, emotion2vec}.
We extend this paradigm to \emph{within-utterance} transition synthesis via mel-space flow blending, which existing corpora~\cite{busso2008iemocap,zhou2022emotional} cannot provide.

\subsection{LLM-Based TTS and Flow Matching}

VALL-E~\cite{valle} pioneered LLM-based zero-shot TTS, followed by Seed-TTS~\cite{seedtts}, NaturalSpeech~3~\cite{naturalspeech3}, MaskGCT~\cite{maskgct}, Llasa~\cite{llasa}, Spark-TTS~\cite{sparktts}, Sticker-TTS~\cite{stickertts}, and GLM-TTS~\cite{Glmtts}.
CosyVoice2~\cite{cosyvoice2}, our backbone, employs supervised semantic tokens with a causal conditional flow matching (CFM)~\cite{lipman2023flow} decoder. CFM has demonstrated strong performance~\cite{matchatts,f5tts,shallowflow}.
Our work extends CFM conditioning to support time-varying emotion.

\subsection{Conditioning Injection in Frozen Models}

Injecting auxiliary signals into frozen models is a growing paradigm.
In audio, FiLM~\cite{film} conditioning appears in WaveGrad~\cite{wavegrad} and DiffWave~\cite{diffwave}; LoRA~\cite{lora} and DoRA~\cite{dora} enable efficient adaptation.
A common thread is the need to control {influence strength}: e.g., ControlNet~\cite{controlnet} uses zero-initialization.
Our direction--magnitude decoupling takes a different route: unlike soft penalties that can be circumvented~\cite{locatello2019challenging}, \emph{LayerNorm + fixed scale} constrains the input-dependent MLP output through re-normalization, so that the injection magnitude remains bounded and close to the speaker-embedding norm.

% ============================================================================
\begin{figure*}[t]
    \centerline{\includegraphics[width=0.95\linewidth]{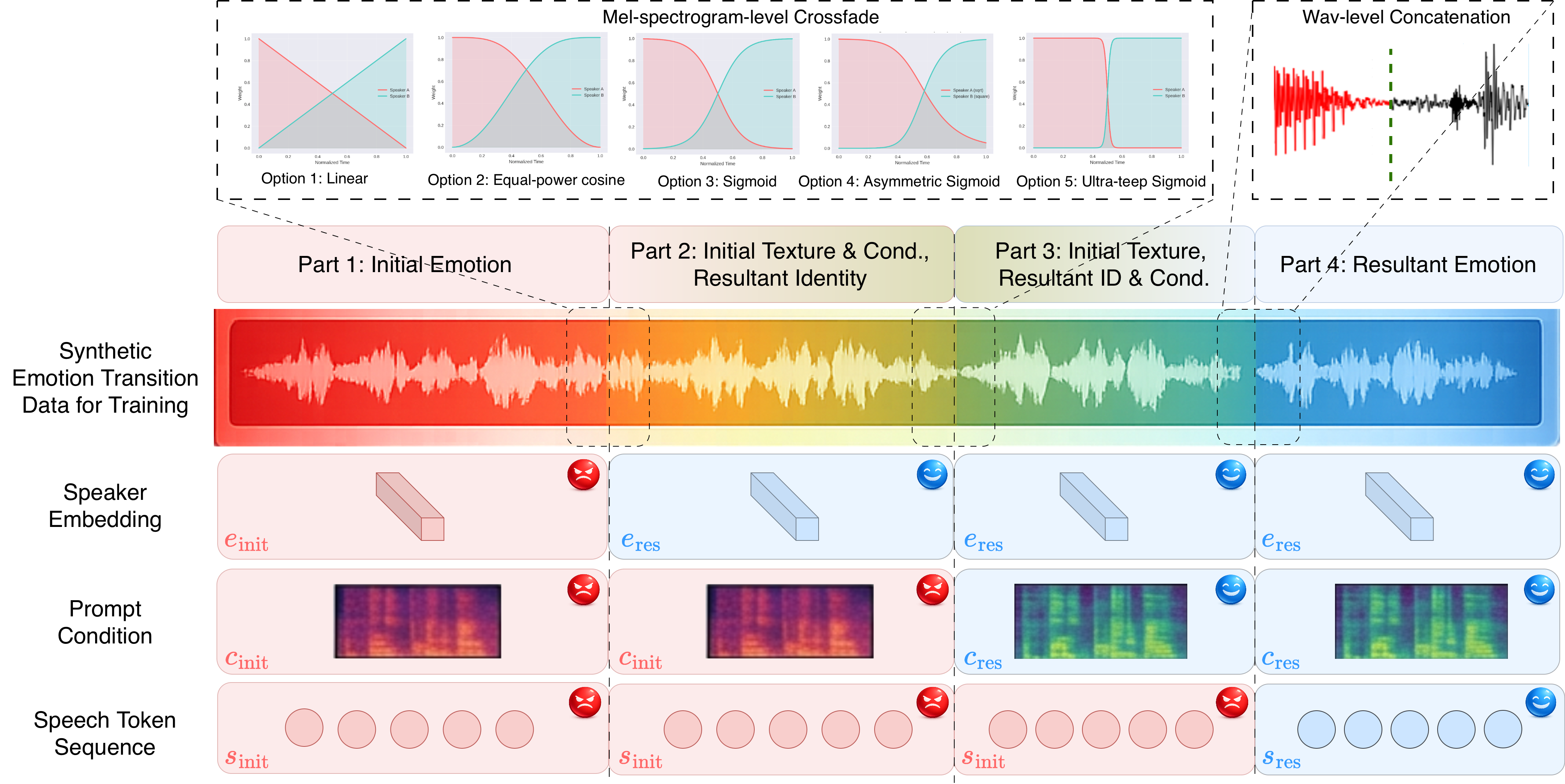}}
    \vspace{-0.05 in}
    \caption{Synthetic emotion transition data generation pipeline. Samples are available on the demo page.}
    \label{fig:systhe}
    % \vspace{-0.10 in}
\end{figure*}

\section{Method}
\label{sec:method}

Research on emotion dynamics shows that affective episodes unfold in phases: from rising intensity to transition or decay~\cite{davidson1998,verduyn2009,kuppens2017emotion}.
Within a single utterance, this manifests as a shift from one affective state to another.
We model this minimal temporal unit as a transition from an \emph{initial emotion} $\bm{v}_\text{init}$ to a \emph{resultant emotion} $\bm{v}_\text{res}$, representing the non-trivial dynamic pattern that existing single-emotion TTS systems cannot express.

\subsection{Problem Formulation}
\label{sec:problem}

Given text $\mathcal{T}$, an initial emotion $\bm{v}_\text{init} = (V_\text{init}, A_\text{init}, D_\text{init}) \in [0,1]^3$, and a resultant emotion $\bm{v}_\text{res} \in [0,1]^3$, we seek to synthesize speech where emotion transitions smoothly from $\bm{v}_\text{init}$ to $\bm{v}_\text{res}$ over the utterance duration.
Framed in terms of emotion dynamics~\cite{davidson1998,kuppens2017emotion}, the system must model the {temporal transition} of an emotional episode, not merely its static endpoint.

We adopt the VAD dimensional model~\cite{russell1980,mehrabian1996} for three reasons:
(a)~it subsumes discrete emotion categories as specific points in the continuous space~\cite{posner2005};
(b)~it naturally supports interpolation for constructing smooth transitions, mirroring gradual emotion shifts in emotion dynamics research~\cite{verduyn2009,cowen2017self};
(c)~its axes have well-established acoustic correlates~\cite{scherer2003,banse1996}, ensuring that changes in VAD space correspond to perceptually meaningful acoustic variations.

\subsection{Synthetic Emotion Transition Data}
\label{sec:data}

\subsubsection{Motivation}
Existing emotional speech corpora are predominantly annotated with single emotion labels~\cite{busso2008iemocap,zhou2022emotional}. Eliciting natural within-utterance transitions from actors requires precisely controlling onset timing, direction, and intensity while maintaining natural prosody and content. These are demands that professional actors find challenging. This data bottleneck has limited prior work on intra-utterance emotion dynamics.

We circumvent this limitation by generating \emph{synthetic emotion transition data} from a pretrained zero-shot TTS system.
The key insight is that flow-based decoders condition on speaker/emotion embeddings {independently of linguistic content}: the same speech token sequence can be decoded under different emotion conditionings. This produces mel spectrograms that are temporally aligned but differ in emotion.
This content-invariance property enables smooth blending in the mel spectrogram space without temporal misalignment.

\subsubsection{Multi-Pass Flow Blending}
\label{sec:multipass}

For a given text and emotion pair (initial, resultant), the pipeline proceeds as follows:

\textbf{(1) Shared token generation.} The LLM generates a single speech token sequence $\bm{s}_\text{init} = [s_1, \ldots, s_M]$ conditioned on the initial emotion. This sequence encodes the prosodic structure and is shared across all subsequent flow passes, ensuring frame-level temporal alignment.

\textbf{(2) Multi-pass flow decoding.} Three independent flow passes produce emotion-specific mel spectrograms (Fig.~\ref{fig:systhe}, Part 1 to Part 3):
\begin{align}
    \text{mel}_\text{init} &= \text{Flow}(\bm{s}_\text{init},\; \bm{e}_\text{init},\; \bm{c}_\text{init}) \\
    \text{mel}_\text{tra} &= \text{Flow}(\bm{s}_\text{init},\; \bm{e}_\text{res},\; \bm{c}_\text{init}) \\
    \text{mel}_\text{res} &= \text{Flow}(\bm{s}_\text{init},\; \bm{e}_\text{res},\; \bm{c}_\text{res})
\end{align}
where $\bm{e}$ denotes the speaker embedding and $\bm{c}$ the prompt condition, both derived from the target speaker.
The intermediate $\text{mel}_\text{tra}$ uses the initial emotion's masked mel spectrogram context but the resultant emotion's speaker embedding, forming a transition bridge between the two emotional states.

\textbf{(3) Sigmoid crossfade blending.} The three mel spectrograms are blended via sigmoid crossfades:
\begin{equation}
    \sigma(t) = \frac{1}{1 + e^{-\kappa(t - t_c)}}
\label{eq:sigmoid}
\end{equation}
with steepness $\kappa$ and center $t_c$, providing near-binary blending at the edges (preserving pure emotion regions) with a smooth crossfade in the center, avoiding audible artifacts from linear blending or hard concatenation.
The choice of the crossfade curve among several candidates, as shown in the upper part of Fig.~\ref{fig:systhe}, is analyzed in Appendix~\ref{sec:crossfade_analysis}.
After vocoding, a fourth part appends a pure resultant-emotion utterance (Part~4 in Fig.~\ref{fig:systhe}).

\textbf{(4) Quality filtering.} Whisper ASR verifies content integrity; samples exceeding CER threshold $\tau_\text{CER}$ or WER threshold $\tau_\text{WER}$ are discarded, as the blending process can occasionally produce phonetically ambiguous regions near transition boundaries.

\subsubsection{Advantages over Recordings}
This pipeline offers three advantages. First, it produces {arbitrary emotion pairs}: any combination of available discrete emotions can be synthesized. Second, transition timing and shape are {precisely controlled} via segment ratios and crossfade parameters, providing consistent supervision. Third, it is {scalable}: multi-GPU parallel synthesis generates thousands of transition samples per hour, whereas studio recording of emotion transitions requires extensive actor preparation and multiple takes.

\begin{figure*}[t]
    \centerline{\includegraphics[width=0.99\textwidth]{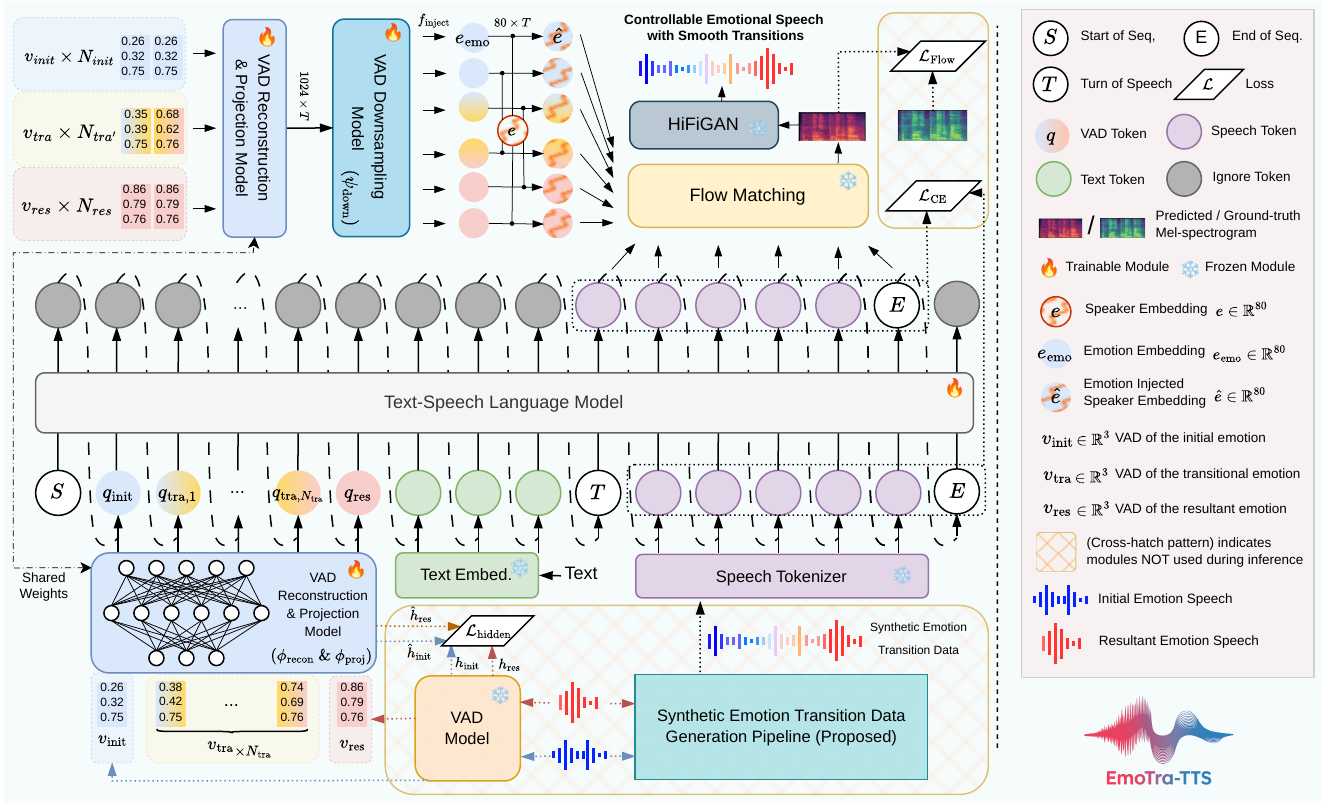}}
    \vspace{-0.08 in}
    \caption{Overview of the EmoTra-TTS. {Lower part}: LLM stage with temporal VAD tokens (\S\ref{sec:llm}). {Upper part}: Flow decoder with frame-level emotion injection via direction--magnitude decoupled injection (\S\ref{sec:flow}).}
    \label{fig:architecture}
    \vspace{-0.12 in}
\end{figure*}

\subsubsection{VAD Annotation}
VAD values for the initial and resultant emotions are obtained by annotating their reference speech with a pretrained wav2vec2-based speech emotion recognition model~\cite{wagner2023dawn}.

\subsection{LLM with Temporal VAD Tokens}
\label{sec:llm}

\subsubsection{VAD Projection}
A projection network $\phi_\text{proj}: \mathbb{R}^3 \rightarrow \mathbb{R}^{d_\text{LLM}}$ maps VAD vectors to the LLM embedding space:
\begin{equation}
    \phi_\text{proj}(\bm{v}) = W_2 \cdot \text{ReLU}(\text{LN}(W_1 \bm{v} + b_1)) + b_2
\end{equation}
where $W_1 \in \mathbb{R}^{d_h \times 3}$, $W_2 \in \mathbb{R}^{d_\text{LLM} \times d_h}$, and LN denotes LayerNorm.
The intermediate normalization stabilizes training given the large dimensionality expansion from 3 to $d_\text{LLM}$, following best practices for cross-modal projection in multimodal learning~\cite{li2023blip2}.

\subsubsection{Temporal Token Design}
A key design question is how to encode a continuous VAD transition in a discrete token sequence.
A single VAD embedding provides only a static global label, while using only the two endpoints $\bm{v}_\text{init}$ and $\bm{v}_\text{res}$ leaves the transition trajectory implicit. In our preliminary experiments, this produced perceptibly abrupt transitions. We therefore uniformly sample $N_\text{tra}$ intermediate points along the line segment between them:
\begin{equation}
    \bm{q}_{\text{tra}, k} = \phi_{\text{proj}}\big((1-\alpha_k)\bm{v}_{\text{init}} + \alpha_k\bm{v}_{\text{res}}\big),
    \alpha_k = \frac{k}{N_{\text{tra}}+1}
    \label{eq:vad_tokens}
\end{equation}
for $k=1, 2, \ldots,N_\text{tra}$.
Together with the endpoint tokens $\bm{q}_\text{init}$ and $\bm{q}_\text{res}$, the $N_\text{tra}+2$ VAD tokens are prepended to the text tokens.
This balances transition resolution with avoiding excessive conditioning that may dilute the LLM’s attention to text.

\subsubsection{Training Objective}
The input sequence is $[\text{SOS}][\bm{q}_\text{init}\; \bm{q}_{\text{tra}, 1} \cdots $ $\bm{q}_{\text{tra}, N_\text{tra}}\; \bm{q}_\text{res}][\text{text}][\text{task}][\text{speech}][\text{EOS}]$, with the loss:
\begin{equation}
    \mathcal{L}_\text{LLM} = \mathcal{L}_\text{CE} + \lambda \cdot \mathcal{L}_\text{hidden}
\label{eq:llm_loss}
\end{equation}
where $\mathcal{L}_\text{CE}$ is the cross-entropy loss on speech tokens.
The auxiliary hidden-state reconstruction loss provides {perceptual grounding}. Without it, $\phi_\text{proj}$ could map VAD vectors to arbitrary embedding regions that are numerically distinct but acoustically meaningless.
\begin{align}
    \hat{\bm{h}}_\text{init} &= \phi_\text{recon}(\bm{q}_\text{init}), \quad
    \hat{\bm{h}}_\text{res} = \phi_\text{recon}(\bm{q}_\text{res})
\end{align}
\begin{equation}
    \mathcal{L}_\text{hidden} = \|\hat{\bm{h}}_\text{init} - \bm{h}_\text{init}\|_1 + \|\hat{\bm{h}}_\text{res} - \bm{h}_\text{res}\|_1
\end{equation}
where $\phi_\text{recon}: \mathbb{R}^{d_\text{LLM}} \rightarrow \mathbb{R}^{d_\text{SER}}$ maps to the wav2vec2 feature space~\cite{wagner2023dawn}, anchoring VAD embeddings to perceptually meaningful prosodic and affective representations rather than arbitrary embedding space regions.

\subsection{Flow Decoder with Frame-Level Emotion Injection}
\label{sec:flow}

\subsubsection{Emotion via the Speaker Pathway}
\label{sec:speaker_pathway}

The backbone flow decoder receives four-channel input $[\bm{x}_t, \bm{\mu}, \bm{e}, \bm{c}]$~\cite{cosyvoice2}, each $\in \mathbb{R}^{d_\text{mel}}$, where $\bm{\mu}$ encodes linguistic content and $\bm{e}$ encodes speaker identity.
We inject emotion through the speaker channel based on two observations.
First, speaker identity and emotion both modulate \emph{how speech sounds} without changing \emph{what is said}; they share the same conditioning role relative to content~\cite{gradstylspeech}.
Second, additive injection through an existing channel requires {no architectural modification} to the pretrained decoder, following the minimal intervention principle that has proven effective in adapter-based approaches~\cite{controlnet,hypertts}:
\begin{equation}
    \hat{\bm{e}}(t) = \bm{e} + \bm{e}_\text{emo}(t)
\label{eq:injection}
\end{equation}
This extends the static 2D speaker tensor to a time-varying 3D tensor $\hat{\bm{e}} \in \mathbb{R}^{B \times d_\text{mel} \times T}$.

\subsubsection{Frame-Level VAD Interpolation}
The frame-level VAD $\bm{v}(t)$ is constructed via piecewise linear interpolation: $\alpha(t) = 0$ for $t < r_\text{init} \cdot T$, linearly interpolated during $[r_\text{init} \cdot T,\, (r_\text{init}+r_\text{tra}) \cdot T)$, and $\alpha(t)=1$ thereafter.
This mirrors the temporal structure of the synthetic training data (\S\ref{sec:data}), ensuring consistency between training supervision and inference-time conditioning.

\subsubsection{VAD Embedding Pipeline}
The emotion embedding reuses frozen LLM-stage modules for cross-stage consistency:
\begin{equation}
    \bm{e}_\text{emo}(t) = f_\text{inject}\big(\psi_\text{down}(\phi_\text{recon}(\phi_\text{proj}(\bm{v}(t))))\big)
\label{eq:pipeline}
\end{equation}
where $\phi_\text{proj}$ and $\phi_\text{recon}$ are frozen from Stage~1, and $\psi_\text{down}: \mathbb{R}^{d_\text{SER}} \rightarrow \mathbb{R}^{d_\text{mel}}$ is a trainable MLP with zero-initialized output layer.
Freezing and reusing VAD modules ensures the same VAD vector produces identical intermediate representations at both stages, preserving the perceptual grounding learned during Stage~1 without additional supervision.

\subsubsection{Direction--Magnitude Decoupled Injection}
\label{sec:method_injection}

The choice of $f_\text{inject}$ is the central design challenge.
An MLP conditioning vector entangles two semantically distinct factors: a \emph{direction} (specifying \emph{which} emotion) and a \emph{magnitude} (specifying \emph{how strongly} it modulates the decoder).
Our ablation (\S\ref{sec:results_ablation} and Appendix~\ref{sec:failure_analysis}) shows that unconstrained training allows MLP output norms to exceed the speaker embedding norm, causing content degradation.
Learnable constraints (e.g., FiLM~\cite{film}) fail because the MLP co-adapts to circumvent them, paralleling the well-documented insufficiency of soft penalties for disentanglement~\cite{locatello2019challenging}.
The flow matching loss, averaging over all mel frames, provides no gradient signal that explicitly constrains the emotion norm, creating exactly the conditions under which soft constraints fail.

We enforce a \emph{hard} separation through architectural constraints:
\begin{equation}
    f_\text{inject}(\bm{x}) = \epsilon \cdot \text{LayerNorm}(\bm{x})
\label{eq:direction_magnitude}
\end{equation}
% LayerNorm locks the output norm at $\approx \sqrt{d_\text{mel}}$ regardless of MLP weights~\cite{ba2016layer}, so the MLP learns emotion \emph{directions} freely but cannot inflate \emph{magnitude}.
LayerNorm re-normalizes the MLP output, so the MLP learns the emotion direction while the injection magnitude stays controlled; the trainable affine parameters ($\gamma$, $\beta$) are few (160) and identity-initialized, and the measured post-training norm is reported in \S\ref{sec:ablation_epsilon}.
The non-trainable scalar $\epsilon$ independently controls injection strength, with no gradient path connecting the two factors.
Zero-initialized output layers ensure $\bm{e}_\text{emo}(t) = \bm{0}$ at training start~\cite{controlnet}.
This design mirrors the direction--magnitude decomposition in DoRA~\cite{dora}, which shows that fine-tuning primarily changes weight \emph{directions}.
The complete injection is:
\begin{equation}
    \hat{\bm{e}}(t) = \bm{e} + \epsilon \cdot \text{LN}\big(\psi_\text{down}(\phi_\text{recon}(\phi_\text{proj}(\bm{v}(t))))\big)
\label{eq:full_injection}
\end{equation}
All decoder parameters are frozen; only $\psi_\text{down}$ and LayerNorm affine parameters are trainable.
Speech tokens are pre-computed from Stage~1, fully decoupling the two training stages.

\subsubsection{Flow Matching Training Objective}
\label{sec:flow_loss}

The flow decoder is trained with a conditional flow matching (CFM)~\cite{lipman2023flow} objective.
Given target mel spectrogram $\bm{x}_1 \in \mathbb{R}^{d_\text{mel} \times T}$, noise $\bm{z} \sim \mathcal{N}(\bm{0}, \bm{I})$, and time $t \sim \mathcal{U}(0,1)$, we construct the interpolated sample $\bm{x}_t = (1 - (1-\sigma_\text{min})t)\,\bm{z} + t\,\bm{x}_1$ and ground-truth velocity $\bm{u} = \bm{x}_1 - (1-\sigma_\text{min})\,\bm{z}$, with $\sigma_\text{min} = 10^{-6}$.
The loss minimizes:
\begin{equation}
    \mathcal{L}_\text{Flow} = \mathbb{E}_{t,\,\bm{x}_1,\,\bm{z}}\!\left[\left\| \hat{\bm{u}}_\theta\!\left(\bm{x}_t,\, t,\, \bm{\mu},\, \hat{\bm{e}}(t)\right) - \bm{u} \right\|^2 \right]
\label{eq:flow_loss}
\end{equation}
where $\hat{\bm{u}}_\theta$ is the velocity estimator and $\hat{\bm{e}}(t)$ is from Eq.~\eqref{eq:full_injection}.
Classifier-free guidance~\cite{ho2022classifier} is applied by randomly dropping conditioning with probability $p_\text{cfg}=0.2$.

% ============================================================================
\section{Experimental Setup}
\label{sec:experiments}

\textbf{Base model.} CosyVoice2-0.5B~\cite{cosyvoice2}.

\textbf{Source corpus.} EmoVoice-DB~\cite{EmoVoice}, containing five speakers and seven emotion categories (\emph{neutral, happy, sad, angry, fearful, disgusted, surprised}).
We apply {cross-validated filtering}: a pretrained wav2vec2-based VAD predictor~\cite{wagner2023dawn}\footnote{{\url{https://github.com/audeering/w2v2-how-to}}} estimates continuous scores, and only samples falling within literature-informed acceptance ranges~\cite{russell1980,mehrabian1996,fontaine2007} (Table~\ref{tab:vad_ranges} in Appendix~\ref{sec:vad_ranges}) are retained, removing mislabeled or acoustically ambiguous samples.

\textbf{Synthetic data.}
Emotion transition pairs are generated via multi-pass flow blending (\S\ref{sec:data}), covering all emotion pair combinations ($\sim$100K utterances).
Quality filtering uses Whisper large-v3~\cite{whisper} with $\tau_\text{CER} = 0.10$; minimum VAD change threshold $\delta_\text{vad}=0.35$ (Chebyshev distance).
Data split: 90/10 train/validation.

\textbf{Training.}
Stage~1 (LLM) fine-tunes the VAD projection and reconstruction modules; Stage~2 (Flow) freezes all decoder parameters and trains only the injection MLP and LayerNorm affine ($\sim$280K parameters).
Fixed emotion scale $\epsilon\!=\!0.07$ yields an effective norm matched to the speaker embedding norm ($\|\bm{e}\|_2 \approx 0.622$).
Full hyperparameters are provided in Appendix~\ref{sec:Hyperparameters}.

\textbf{Evaluation.}
We conduct a blind listening test with a professional panel of 21 raters.
To enable a more fine-grained and independent evaluation of different aspects, we decompose MOS into three metrics:
\textbf{MOS-Qua}, for speech naturalness and quality (ignoring emotional correctness); \textbf{MOS-Emo} for emotion adequacy; and \textbf{MOS-Tra} for emotion transition smoothness. 
We compute \textbf{speaker similarity} \textbf{(SIM)}\footnote{{\url{https://github.com/BytedanceSpeech/seed-tts-eval}}}, with the reference defined as the average embedding of 200 samples covering all emotions.
\textbf{Word error rate (WER)} is evaluated by Whisper large-v3\footnote{\url{https://huggingface.co/openai/whisper-large-v3}}~\cite{whisper}. 
We further include two {pairwise preference tests} with partially overlapping rater pools: one against the open-source baselines (11 raters, 160 pairs, $\sim$6$\times$ coverage; \S\ref{sec:Pairwise}, Appendix~\ref{sec:paired_protocol}) and a supplementary one against two commercial systems (12 raters, 80 pairs, $\sim$8$\times$ coverage; \S\ref{sec:commercial_main}, Appendix~\ref{sec:commercial_protocol}); both report \textbf{win rates, 95\% CIs}, and \textbf{inter-rater agreement}.
In addition, we compute the \textbf{Prosodic Jerk Ratio (JR-F0) }in Appendix~\ref{sec:objective} as an {objective metric} for transitions.

% \footnotetext[4]{{}}
% ============================================================================
\section{Results and Analysis}
\label{sec:results}

\subsection{Injection Architecture Comparison}
\label{sec:results_ablation}

\begin{table}[h]
\centering
\caption{Injection architecture comparison.}
\vspace{-0.05 in}
\label{tab:ablation_summary}
\resizebox{\linewidth}{!}{
\begin{tabular}{@{}lcclll@{}}
\toprule
\multirow{2}{*}{\textbf{Variant}} & \multicolumn{3}{c}{\textbf{Quality Evaluation}} & \multicolumn{2}{c}{\textbf{Emotion Evaluation}} \\ \cmidrule(r){2-4} \cmidrule(r){5-6} 
 & MOS-Qua & SIM & WER & MOS-Emo & MOS-Tra \\ 
\midrule
Linear                    & \textbf{3.60}$\pm$0.12 &\textbf{0.707} & \textbf{1.65} &  3.39$\pm$0.15 & 3.33$\pm$0.16\\
\textbf{Dir--Mag (ours)}  & 3.54$\pm$0.13 & 0.701 & 1.77 &  \textbf{3.52}$\pm$0.15 & \textbf{3.63}$\pm$0.17 \\
\bottomrule
\end{tabular}
}
\vspace{-0.05 in}
\end{table}

\begin{table*}[t]
\centering
\caption{Comparison with existing TTS systems on the test set. SIM: speaker similarity. WER: word error rate (\%). We decompose MOS into MOS-Qua, MOS-Emo, and MOS-Tra for a more fine-grained and independent evaluation of speech quality, emotion rendering adequacy, and intra-utterance emotion transition smoothness, respectively. Subjective evaluation results are reported as means with 95\% CIs. Our base system is {CosyVoice2}. }
\vspace{-0.07 in}
\resizebox{0.99\linewidth}{!}{
\begin{tabular}{@{}lccllcc@{}}
\toprule
\multirow{2}{*}{\textbf{Model}} & \multirow{2}{*}{\begin{tabular}[c]{@{}c@{}}\textbf{Emotion}\\\textbf{Conditioning}\end{tabular}} & \multicolumn{3}{c}{\textbf{Quality Evaluation}} & \textbf{Emotion Rendering} & \textbf{Emotion Transition} \\ \cmidrule(r){3-5} \cmidrule(r){6-6} \cmidrule(r){7-7} 
 &  & MOS-Qua$\uparrow$ & SIM$\uparrow$ & WER$\downarrow$ & MOS-Emo$\uparrow$ & MOS-Tra$\uparrow$ \\ 
\midrule
\multicolumn{7}{l}{\textit{(a) Neutral prompt speech; no emotion conditioning}} \\
\hline
\ \ \ \ Qwen3-TTS~\cite{Qwen3TTS} & ---     & \textbf{4.02}$\pm$0.09 & 0.726 & 1.05 & 2.72$\pm$0.17 & 2.40$\pm$0.18 \\ 
\ \ \ \ MOSS-TTS~\cite{mosstts} & ---       & 3.98$\pm$0.09 &\textbf{0.754}  & 1.01 & 2.76$\pm$0.18 & 2.36$\pm$0.17 \\ 
\ \ \ \ CosyVoice2~\cite{cosyvoice2} & --- & 3.87$\pm$0.10 & 0.711 & 1.21 & 2.53$\pm$0.16 & 2.32$\pm$0.17 \\ 
\hline
\multicolumn{7}{l}{\textit{(b) Natural-language emotion instruction (e.g., ``speak happily then sadly'')}} \\
\hline
\ \ \ \ Qwen3-TTS~\cite{Qwen3TTS}      & Instruct & 3.79$\pm$0.10 & N/A  & 1.28 & 3.33$\pm$0.15  & 2.67$\pm$0.19 \\ 
\ \  \ \ MOSS-TTS~\cite{mosstts}       & Instruct & 3.22$\pm$0.15 & N/A  & 2.55 & 2.80$\pm$0.17 & 2.13$\pm$0.16 \\ 
\hline
\multicolumn{7}{l}{\textit{(c) Natural-language emotion instruction with neutral prompt speech}} \\
\hline
\ \  \ \ CosyVoice2~\cite{cosyvoice2} & Instruct & 3.63$\pm$0.12 & 0.634  & 1.60 & 2.64$\pm$0.14 & 2.15$\pm$0.16 \\ 
\ \ \ \  EmoVoice~\cite{EmoVoice}      & Instruct & 3.59$\pm$0.13 & 0.640  & 1.42 & 2.48$\pm$0.15 & 2.31$\pm$0.17 \\
\hline
\multicolumn{7}{l}{\textit{(d) Emotional prompt speech: generate each emotion segment separately, then concatenate}} \\
\hline
\ \  \ \ Qwen3-TTS~\cite{Qwen3TTS}     & Prompt & 3.46$\pm$0.14 & 0.730  & \textbf{0.92} & 3.70$\pm$0.14  & 2.80$\pm$0.18 \\
\ \ \ \  MOSS-TTS~\cite{mosstts}       & Prompt & 2.85$\pm$0.17 & 0.735 & 2.64 & 3.89$\pm$0.12 & 1.94$\pm$0.16 \\ 
\ \  \ \ CosyVoice2~\cite{cosyvoice2} & Prompt & 2.97$\pm$0.15 & 0.703 & 1.83 & 3.82$\pm$0.12 & 2.26$\pm$0.17 \\
\hline
\multicolumn{7}{l}{\textit{(e) Emotional prompt speech with word-level emotion generation control}} \\
\hline
\ \  \ \ WeSCon~\cite{wescon}          & Prompt & 2.80$\pm$0.15 & 0.649  & 2.92 & 3.53$\pm$0.14 & 2.39$\pm$0.17 \\ 
\hline
\multicolumn{7}{l}{\textit{(f) Our synthetic training data: multi-pass flow blending (\S\ref{sec:data}), decoded by base CosyVoice2}} \\
\hline
\ \ \ \ \textbf{Synth.\ Data (Ours)} & Prompt & 3.58$\pm$0.11 & 0.707 & 1.49 & \textbf{4.05}$\pm$0.12 & 3.55$\pm$0.15 \\ 
\hline
\multicolumn{7}{l}{\textit{(g) Proposed: continuous VAD conditioning with smooth intra-utterance transition}} \\
\hline
\ \ \ \ \textbf{EmoTra-TTS (Ours)}                      & VAD & 3.54$\pm$0.13 & 0.701 & 1.77 &  3.52$\pm$0.15 & \textbf{3.63}$\pm$0.17 \\  
\ \ \ \ \ \ \ \ -- w/o flow SFT  & VAD & 3.62$\pm$0.12 & 0.709 & 1.65 & 2.70$\pm$0.17 & 2.58$\pm$0.18 \\
\bottomrule
\end{tabular}
}
\label{tab:comparison}
\vspace{-0.05 in}
\end{table*}

We evaluate seven injection architectures for Stage~2 (Table~\ref{tab:ablation_summary}; full results and failure analysis in Appendix~\ref{sec:failure_analysis}, Table~\ref{tab:ablation_full}).
Linear yields the best speech quality (MOS-Qua 3.60, SIM 0.707, WER 1.65\%) but weaker emotion (MOS-Emo 3.39, MOS-Tra 3.33), suggesting that simple additive injection preserves content but lacks expressiveness.
Our Dir--Mag design structurally decouples emotion direction from injection magnitude, achieving the best emotion performance (MOS-Tra \textbf{3.63}, MOS-Emo \textbf{3.52}) with acceptable speech quality trade-off (MOS-Qua 3.54, SIM 0.701, WER 1.77\%).

\subsection{Comparison with Open-source Systems}
\label{sec:comparison}

Table~\ref{tab:comparison} compares EmoTra-TTS against baselines spanning seven conditioning strategies.
These baselines are all recent open-source systems, including CosyVoice2~\cite{cosyvoice2}, EmoVoice~\cite{EmoVoice}, WeSCon~\cite{wescon}, Qwen3-TTS~\cite{Qwen3TTS}, and MOSS-TTS~\cite{mosstts}, which together reasonably represent the current instruction-following capability of open-source TTS.
Closed-source commercial systems are treated separately: recent benchmarks report the same intra-utterance transition limitation (\S\ref{sec:related}), and we provide a direct comparison against two of them (GPT-4o mini TTS and ElevenLabs~v3) in \S\ref{sec:commercial_main}.

\textbf{Neutral baselines~(a)} achieve the highest quality (up to MOS-Qua 4.02), but produce minimal emotion rendering and transition quality. MOS-Qua primarily reflects speech quality rather than emotional correctness; therefore, high MOS-Qua scores under neutral conditions are expected.

\textbf{Instruction-based systems~(b, c)} show that current TTS systems cannot translate natural-language emotion instructions into precise intra-utterance emotion transitions. The instruction-following capability determines emotion rendering (MOS-Emo ranging from 2.48 to 3.33); adding a neutral prompt~(c) preserves speaker similarity (SIM 0.634--0.640) but anchors the decoder to neutral prosody, further limiting emotion expressiveness.

\textbf{Prompt-based concatenation~(d)} achieves the strongest baseline emotion rendering (up to MOS-Emo 3.89) since each segment is cloned from a matching emotional reference. However, stronger emotional expressiveness increases perceptual discontinuity at boundaries. MOSS-TTS exemplifies this: it achieves the highest baseline MOS-Emo (3.89) yet the lowest MOS-Tra (1.94), as vivid emotion makes boundary artifacts more noticeable.

\textbf{High Failure Cost}. We further quantify this effect. When discrete baselines such as MOSS-TTS produce abrupt emotion shifts (Category d), naturalness drops significantly compared to neutral conditions (Category a), showing that a single failed transition can degrade overall quality. Moreover, this degradation increases with higher emotion expressiveness (MOS-Emo), indicating that failure cost grows with expressiveness. This emphasizes the importance of smooth transition modeling in expressive TTS due to its high perceptual impact.

\textbf{WeSCon~(e)} controls intra-utterance emotion variation in a single decoding pass via word-level prompt conditioning. While this enables finer-grained control, its discrete per-word assignment lacks smooth transition modeling between segments. As a result, it introduces boundary discontinuities, leading to lower MOS-Tra (2.39).

All above systems achieve MOS-Tra $\leq$ 2.80, indicating that controllable smooth intra-utterance emotion transition remains an open problem.

\textbf{Our synthetic data~(f)} achieves the highest emotion rendering performance (MOS-Emo \textbf{4.05}) and a MOS-Tra of 3.55, validating the effectiveness of the synthetic data. While outperforming all baselines, the multi-pass flow blending design substantially reduces synthesis efficiency (see \S\ref{sec:efficiency}), motivating the training of our proposed EmoTra-TTS.

\textbf{EmoTra-TTS~(g)} synthesizes the utterance in a {single decoding pass} with continuous frame-level VAD interpolation, enabling smooth rather than abrupt transitions, yielding the best MOS-Tra (\textbf{3.63}) with competitive results on other metrics. Among systems with explicit emotion rendering (Categories d, e), EmoTra-TTS achieves the highest MOS-Qua, corroborated by an objective metric (Appendix~\ref{sec:objective}) showing lower F0 jerk at transitions.

\subsection{Ablation: Dual-Stage SFT}
\label{sec:ablation_sft}

We compare three configurations in Table~\ref{tab:comparison}: baseline CosyVoice2~(a), LLM-only SFT (``w/o flow SFT'' in group~g), and full EmoTra-TTS~(g).
LLM-only SFT improves emotion performance (MOS-Emo $2.53 \!\rightarrow\! 2.70$, MOS-Tra $2.32 \!\rightarrow\! 2.58$) while preserving speech quality, confirming that temporal VAD tokens provide meaningful prosodic planning signals.
Adding flow SFT yields a further substantial gain (MOS-Emo $2.70 \!\rightarrow\! 3.52$, MOS-Tra $2.58 \!\rightarrow\! 3.63$), demonstrating \emph{complementary} contributions: the LLM plans emotion-appropriate prosody, while the flow stage renders acoustic realization through frame-level VAD modulation.

\subsection{Ablation: Fixed Scale \texorpdfstring{$\epsilon$}{epsilon}}
\label{sec:ablation_epsilon}

\begin{table}[h]
\centering
\caption{Ablation on fixed scale $\epsilon$. The effective emotion norm is $\epsilon \cdot \sqrt{d_\text{mel}}$; the speaker embedding norm is $\|\bm{e}\|_2 \approx 0.622$.}
\label{tab:ablation_epsilon}
\vspace{-0.05 in}
\resizebox{\linewidth}{!}{
\begin{tabular}{@{}cccclcc@{}}
\toprule
\multirow{2}{*}{\textbf{$\epsilon$}} & \multirow{2}{*}{\begin{tabular}[c]{@{}c@{}}\textbf{Eff.}\ \textbf{norm}\end{tabular}} & \multicolumn{3}{c}{\textbf{Quality Evaluation}} & \multicolumn{2}{c}{\textbf{Emotion Evaluation}} \\ \cmidrule(r){3-5} \cmidrule(r){6-7} 
 & & MOS-Qua & SIM & WER & MOS-Emo & MOS-Tra \\ 
\midrule
        0.04 & 0.358            & 3.60$\pm$0.12 & {0.706} & 1.69 & 3.46$\pm$0.15 & 3.31$\pm$0.16 \\
\textbf{0.07} & \textbf{0.626} &  3.54$\pm$0.13 & 0.701 & 1.77 &  3.52$\pm$0.15 & \textbf{3.63}$\pm$0.17 \\
        0.10 & 0.894            & 2.59$\pm$0.18 & 0.491 & 5.30 & 2.34$\pm$0.17 & 1.85$\pm$0.17 \\
\bottomrule
\end{tabular}
}
\vspace{-0.15 in}
\end{table}

The fixed scale $\epsilon$ controls the emotion-to-speaker norm ratio $r = \lVert\bm{e}_\text{emo}\rVert_2 \,/\, \lVert\bm{e}\rVert_2$ ($\lVert\bm{e}\rVert_2 \approx 0.622$), where $r<1$ means speaker identity dominates, $r>1$ means emotion dominates, and $r\approx 1$ is parity.  The results reveal a sensitivity boundary: $\epsilon=0.04$ ($r=0.58$) preserves speech quality (MOS-Qua 3.60) but attenuates emotion; $\epsilon=0.10$ ($r=1.44$) triggers catastrophic degradation as emotion injection overwhelms content and speaker identity. The optimal $\epsilon=0.07$ ($r\approx 1.0$) achieves the best transition quality (MOS-Tra \textbf{3.63}), confirming that parity between emotion and speaker norms defines the safe operating regime.

At the deployed checkpoint ($\epsilon\!=\!0.07$), the \emph{measured} post-training injection norm is $\|\bm{e}_\text{emo}(t)\|_2 \approx 0.667$, about $1.07\times$ the speaker-embedding norm ($\approx 0.622$); the $0.626$ in Table~\ref{tab:ablation_epsilon} is the \emph{nominal} $\epsilon\cdot\sqrt{d_\text{mel}}$.
That the measured norm stays close to the speaker norm confirms that the trainable LayerNorm affine parameters ($\gamma,\beta$; 160 in total, identity-initialized) do not inflate the injection magnitude, keeping it far below the failure regime, where an input-dependent MLP inflates the norm to $\approx 2\times$ (MLP-Add) or $3$--$5\times$ (FiLM) and collapses content (Appendix~\ref{sec:failure_analysis}).
\subsection{Pairwise Preference Evaluation}
\label{sec:Pairwise}
\begin{figure}[t]
    \centerline{\includegraphics[width=0.99\linewidth]{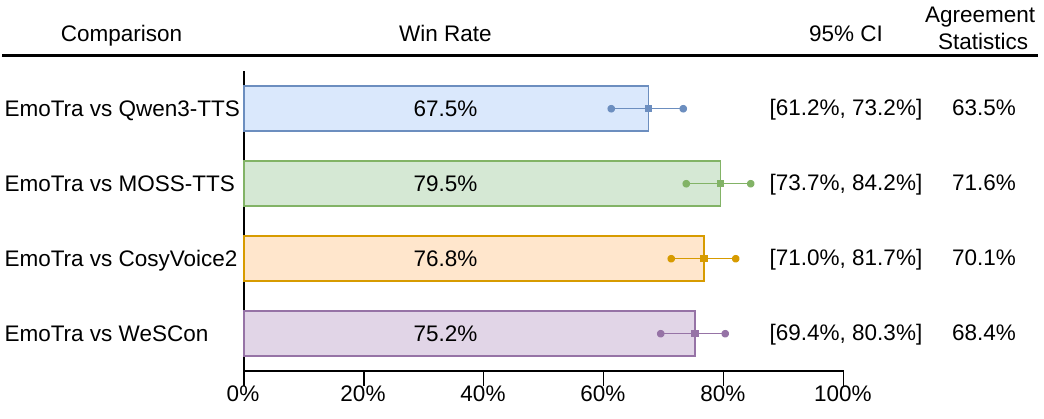}}
    \vspace{-0.05 in}
    \caption{Paired preference of EmoTra-TTS over four baselines in Categories (d) and (e) of Table~\ref{tab:comparison}.}
    \label{fig:pair}
    \vspace{-0.1 in}
\end{figure}
Beyond the per-dimension MOS above, we conduct a paired preference test for an overall A/B comparison. Figure~\ref{fig:pair} shows EmoTra-TTS is preferred over all four baselines with win rates of 67.5\%–79.5\%, and every 95\% CI lower bound exceeds 60\%, well above the 50\% chance level. Inter-rater agreement of 63.5\%–71.6\% supports these preferences.

\subsection{Comparison with Commercial Systems}
\label{sec:commercial_main}
As a supplementary study added during the revision period, we additionally compare EmoTra-TTS against two closed-source commercial systems spanning the two widely adopted commercial control paradigms.
This study uses a partially overlapping but not identical rater pool and set of pairs; it therefore constitutes a valid EmoTra-vs.-commercial comparison but is \emph{not} directly comparable to the paired results in \S\ref{sec:Pairwise} and Figure~\ref{fig:pair}. The full protocol is provided in Appendix~\ref{sec:commercial_protocol}.

\textbf{ElevenLabs~v3 (latest)}, from one of the most widely used commercial TTS providers, controls emotion through discrete audio tags rather than natural-language prompts.
\textbf{GPT-4o mini TTS}, from OpenAI's GPT-4o audio family, is controlled via a natural-language instruction prompt; as it is the same family used to synthesize our source corpus EmoVoice-DB~\cite{EmoVoice}, the comparison is especially stringent.
The two systems thus span the natural-language instruction setting (GPT-4o mini TTS) and the discrete-tag setting (ElevenLabs~v3) available in commercial products; configuration and prompting details are given in Appendix~\ref{sec:commercial_protocol}.

\begin{table}[t]
\centering
\caption{Comparison with commercial instruction-following systems. Values are the percentage of pairs in which EmoTra-TTS is preferred (win rate), with 95\% CIs. ``Emo.\ \& Tra.'' evaluates emotion rendering and transition naturalness; ``Overall'' is overall preference. ``Agr.'' denotes inter-rater agreement.}
\label{tab:commercial}
\vspace{-0.05 in}
\resizebox{\linewidth}{!}{
\begin{tabular}{@{}lcccccc@{}}
\toprule
\multirow{2}{*}{\textbf{EmoTra-TTS vs.}} & \multicolumn{3}{c}{\textbf{Emo.\ \& Tra.}} & \multicolumn{3}{c}{\textbf{Overall}} \\
\cmidrule(r){2-4}\cmidrule(r){5-7}
 & Win & 95\% CI & Agr. & Win & 95\% CI & Agr. \\
\midrule
GPT-4o mini TTS & 74.5\% & [69.5, 78.9] & 67.7\% & 72.0\% & [66.9, 76.6] & 63.8\% \\
ElevenLabs v3   & 70.5\% & [65.2, 75.3] & 63.5\% & 64.4\% & [59.0, 69.5] & 60.7\% \\
\bottomrule
\end{tabular}
}
\vspace{-0.1 in}
\end{table}

As shown in Table~\ref{tab:commercial}, EmoTra-TTS is preferred over both commercial systems in overall preference (72.0\% and 64.4\%), with larger margins on emotion accuracy and transition naturalness (74.5\% and 70.5\%); every 95\% CI lower bound exceeds the 50\% chance level.

\subsection{Inference Efficiency}
\label{sec:efficiency}

\begin{table}[h]
\centering
\vspace{-0.05 in}
\caption{Inference efficiency comparison. RTF: real-time factor. Mean ± std reported. Mem.: GPU memory.}
\label{tab:speed}
\vspace{-0.05 in}
\resizebox{\linewidth}{!}{%
\begin{tabular}{@{}lccccc@{}}
\toprule
\textbf{System} & \textbf{RTF}$\downarrow$ & \textbf{Latency (s)}$\downarrow$ & \textbf{Params} & \textbf{$\Delta$Params} & \textbf{Mem.}\\
\midrule
CosyVoice2 & 0.270$\pm$.022 & 1.75$\pm$.39 & 639M & --  & 3.04 GB\\
Synth.\ Data & 0.397$\pm$.045 & 2.92$\pm$.47 & 639M & +0.00\%  & 5.09 GB \\
\textbf{EmoTra-TTS} & 0.244$\pm$.010 & 1.74$\pm$.42 & 642M & +0.43\% & 3.05 GB  \\
\bottomrule
\end{tabular}%
}
\vspace{-0.05 in}
\end{table}

EmoTra-TTS's latency is close to the baseline, indicating negligible overhead from the additional 0.43\% parameters.  The lower RTF results from longer generated audio rather than faster synthesis. The synthetic data pipeline requires $4\times$ flow inference and is intended for offline generation. See Appendix~\ref{sec:statis_infer} for test set details.

% ============================================================================
\section{Conclusion}
\label{sec:conclusion}

Emotion dynamics research has long established that affect unfolds continuously over time, yet explicit, controllable modeling of the temporal trajectory of emotion within a single utterance is still only partially addressed by existing systems.
EmoTra-TTS bridges this gap through multi-pass flow blending for synthetic transition data, dual-stage VAD conditioning for complementary prosodic and acoustic control, and direction--magnitude decoupled injection that constrains the injection magnitude to alleviate the content-expressiveness trade-off.
The system adds only $+0.43\%$ parameters with no latency overhead and achieves the best perceptual scores on emotion transition smoothness among all evaluated systems, with 30\%–87\% relative improvement over multiple SOTA TTS baselines and commercial systems, and 64.4\%–79.5\% win rates in pairwise preference tests.
We hope this work offers a useful step toward modeling emotion as a \emph{temporal process} rather than a static attribute, and encourages further exploration of affective speech synthesis.

\section*{Limitations}

Our work has several limitations:

First, the current experiments are conducted on EmoVoice-DB, which provides limited speaker diversity and language coverage. While we validate EmoTra-TTS on multiple speakers from EmoVoice-DB, a separate model is trained per voice, and zero-shot generalization to arbitrary unseen speakers is not yet demonstrated. This limitation, however, mainly concerns speaker and linguistic generalization rather than the proposed intra-utterance emotion transition mechanism itself. Extending the framework to larger multilingual multi-speaker emotional corpora and incorporating prompt-based speaker conditioning remain important directions for future work.

Second, emotion control stability can be inconsistent for subtle VAD differences, likely improvable with more diverse training data.

Third, the current piecewise linear VAD interpolation provides only a first-order approximation of natural emotion dynamics~\cite{davidson1998}. More complex nonlinear trajectories may better capture realistic affective evolution within an utterance.

Fourth, a formal characterization of conditioning tolerance boundaries in frozen generative models remains an open research problem.

\section*{Ethical Statement}

\textbf{Human annotation and fair compensation.}
All evaluators involved in the blind listening test were either formally employed researchers or graduate students supported by institutional scholarships.
The blind listening test work was compensated in accordance with local labor regulations and institutional guidelines, consistent with ACL requirements regarding fair treatment and remuneration of human participants.
No personally identifiable data were collected from evaluators, and the task involved only subjective quality ratings of AI-synthesized speech; our institution classifies this as minimal-risk research exempt from formal ethics board review.

\textbf{Data privacy and consent.}
All training data are synthesized from EmoVoice-DB~\cite{EmoVoice}, a publicly available emotional speech corpus released for research purposes. EmoVoice-DB is itself entirely AI-generated and contains no human recordings.
The synthetic emotion transition data are generated entirely from this public dataset via our multi-pass flow blending pipeline; no private, user-uploaded, or personally identifiable data are used at any stage.
The released data and model checkpoints do not contain any personally identifiable information or sensitive user data.

\textbf{Licensing and responsible use.}
The complete code, including both the synthetic data pipeline and model training pipelines, will be released under an open-source license upon acceptance, restricted to non-commercial academic research.
We acknowledge that controllable emotional speech synthesis carries potential risks of misuse, including generating deceptive or manipulative audio content.
We emphasize that EmoTra-TTS is intended as a research contribution to advance affectively dynamic speech synthesis, and we encourage responsible use with appropriate human oversight in any downstream application.

\textbf{Usage of AI assistants.} AI language models were used in two capacities: (1)~generating test set sentences with emotion transition annotations (Appendix~\ref{sec:test_prompt}), and (2)~language polishing during paper writing. All experimental design, model development, analysis, and scientific conclusions were made by the authors.

\bibliography{custom}

\appendix

\begin{figure}[h]
    \centering
    \includegraphics[width=\linewidth]{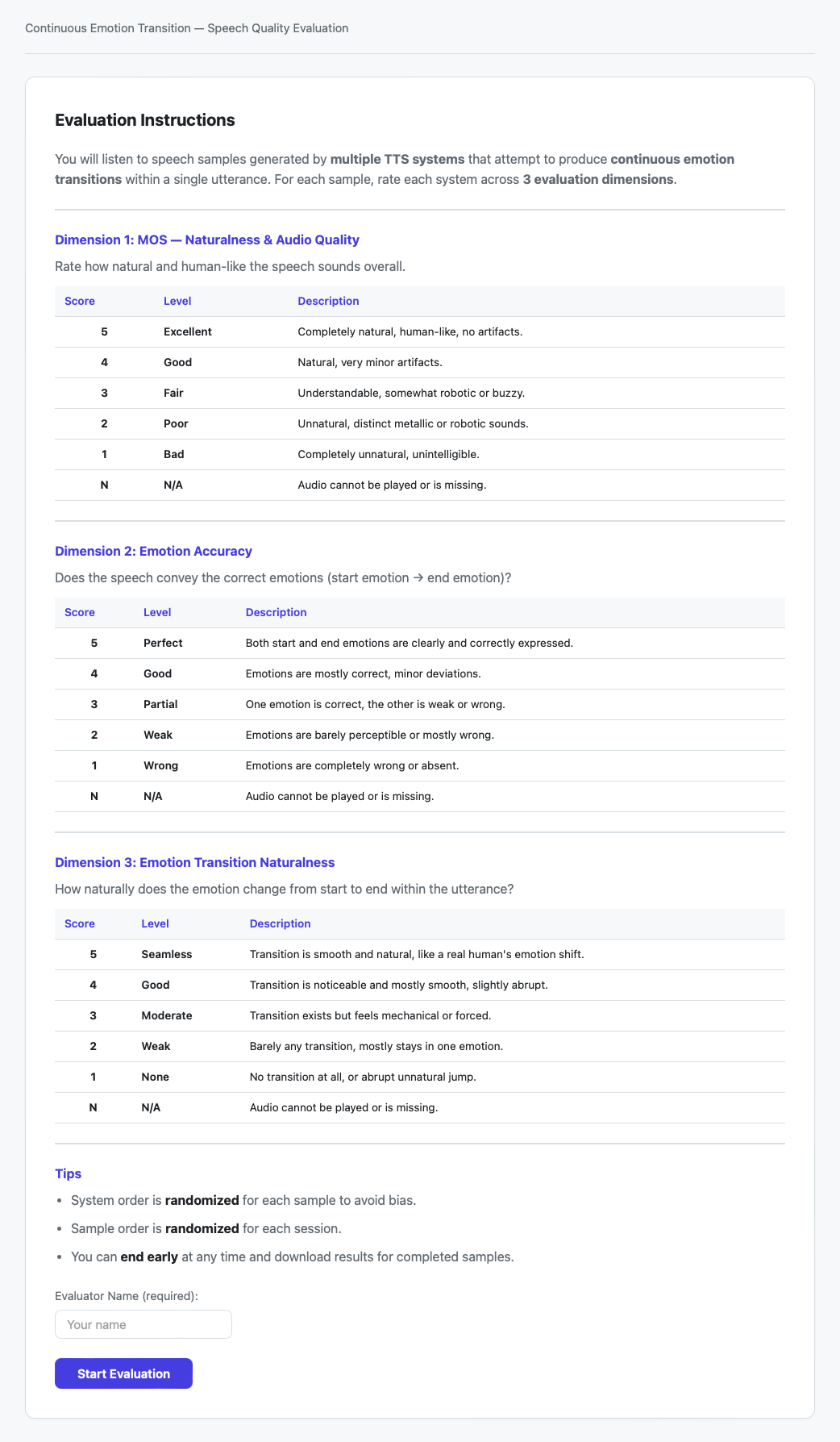}
    \vspace{-0.2 in}
    \caption{Rating criteria for `Naturalness \& Audio Quality' (MOS-Qua), `Emotion Accuracy' (MOS-Emo), and `Emotion Transition Naturalness' (MOS-Tra).}
    \label{fig:rating_criteria}
    \vspace{-0.1 in}
\end{figure}

\begin{figure}[t]
    \centering
    \includegraphics[width=\linewidth]{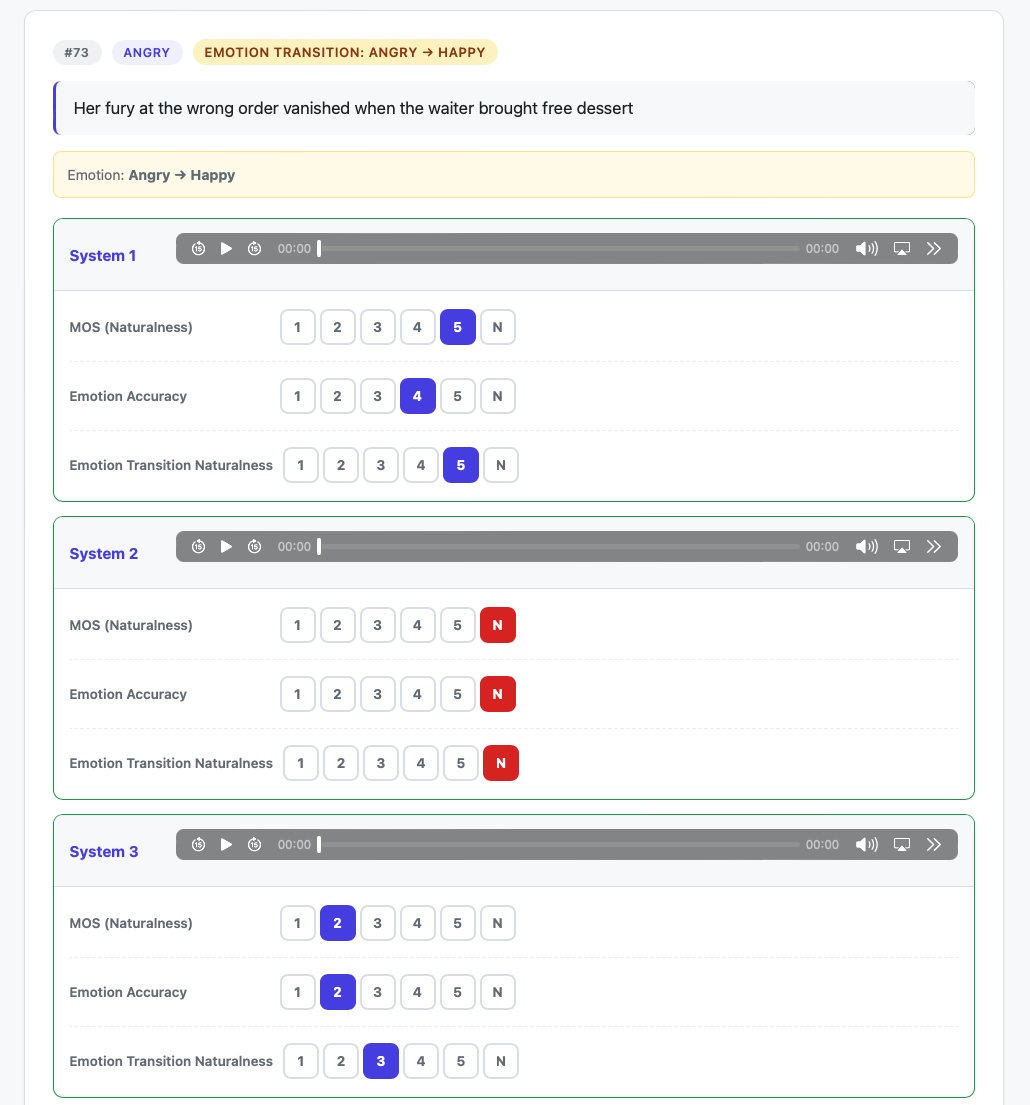}
    \vspace{-0.2 in}
    \caption{The blind test evaluation interface. Evaluators listen to each sample and rate `MOS Naturalness' (MOS-Qua), `Emotion Accuracy' (MOS-Emo), and `Emotion Transition Naturalness' (MOS-Tra) on 1--5 Likert scales. System identities are hidden, and the system order is randomly shuffled for each sample.}
    \label{fig:eval_interface}
    \vspace{-0.10 in}
\end{figure}

\section{Training Details}
\label{sec:Hyperparameters}

\textbf{Synthetic data generation:}
Crossfade steepness $\kappa=12$; segment ratios $r_\text{init}/r_\text{tra}/r_\text{res} = 60\%/15\%/25\%$; crossfade duration $\min(0.4\text{s}, 0.1 \times T_\text{dur})$.
ASR quality filter: Whisper large-v3 with CER threshold $\tau_\text{CER}=0.10$.

\textbf{Stage~1 training (LLM):}
Adam optimizer, learning rate $= 1.0 \times 10^{-5}$, gradient clipping at $5.0$.
$N_\text{tra}=3$ intermediate VAD tokens ($N_\text{tra}+2=5$ total), $d_h=256$, $d_\text{LLM}=896$, $d_\text{SER}=1024$, $\lambda=0.1$.

\textbf{Stage~2 training (Flow):}
All decoder parameters frozen.
Adam optimizer, gradient accumulation factor of 2.
Segment ratios: $r_\text{init}=0.4$, $r_\text{tra}=0.3$, $r_\text{res}=0.3$.
Fixed emotion scale $\epsilon=0.07$, yielding effective norm $\approx \epsilon \cdot \sqrt{d_\text{mel}} = 0.07 \times \sqrt{80} \approx 0.626$, matched to the observed speaker embedding norm ($\|\bm{e}\|_2 \approx 0.622$).
$\psi_\text{down}$: $1024 \rightarrow 256 \rightarrow \text{ReLU} \rightarrow 80$ (last layer zero-initialized).
LayerNorm: $d_\text{mel}=80$.

\textbf{Evaluation cost.}
The two commercial systems used for the paired comparison in \S\ref{sec:commercial_main} incurred a total cost of USD~11 (ElevenLabs subscription: USD~6; GPT-4o mini TTS API: USD~5).

\section{Evaluation Protocol}
\label{sec:appen_eval}

We conduct blind listening evaluations using a professional paid blind-test team with established quality control procedures. In total, 21 raters participated across all evaluations, including researchers and non-technical listeners.

\subsection{MOS-based Evaluation Protocol}
\label{sec:MOSProtocol}
For the MOS-based evaluation, 10 evaluators (3 researchers and 7 non-technical listeners) participated to capture both expert and general-listener perspectives. Each system contains 200 utterances, and each evaluator rated about 25 randomly sampled utterances per system, resulting in approximately 250 ratings per system and about 25\% repeated coverage. 

\begin{figure}[t]
    \centering
    \includegraphics[width=\linewidth]{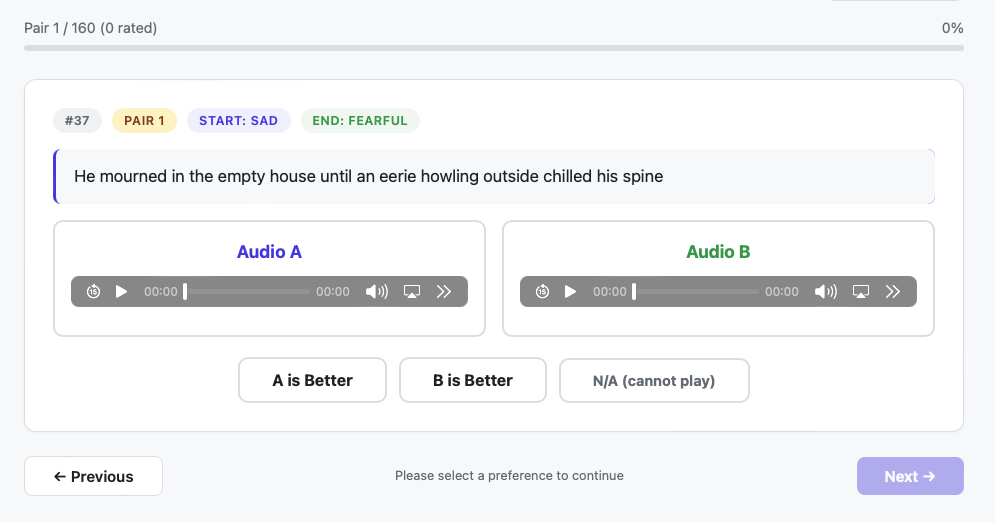}
    \vspace{-0.2 in}
    \caption{Screenshot of the paired preference evaluation interface. System identities are hidden, and the system order is randomly shuffled for each sample.}
    \label{fig:eval_interface2}
    \vspace{-0.1 in}
\end{figure}

These test sample texts and target emotions are generated by {ChatGPT-5.2}~\footnote{\url{https://help.openai.com/en/articles/6825453-chatgpt-release-notes}} using a structured prompt (Appendix~\ref{sec:test_prompt}) that enforces event-driven narratives, balanced emotion-pair coverage, and VAD annotations within tight per-category ranges.
The resulting test set is independent of the EmoVoice-DB~\cite{EmoVoice} training set.

Notably, we prioritize the \textbf{quality} of the blind test over its \textbf{quantity}. With a carefully selected panel, we emphasize careful, attentive ratings rather than maximizing raw throughput, while still achieving a high overall completion rate of \textasciitilde 94\%. All samples are randomly shuffled and system identities are hidden from evaluators.

Three metrics are collected on a 1--5 Likert scale:
\begin{itemize}
    \item \textbf{MOS-Qua}: speech naturalness and quality, without considering emotional correctness.
    \item \textbf{MOS-Emo}: adequacy of emotion rendering relative to the target specification, evaluating whether both the initial and resultant emotions are correctly expressed.
    \item \textbf{MOS-Tra}: smoothness and naturalness of intra-utterance emotion transition, assessing whether the shift between emotions sounds gradual and natural rather than abrupt or discontinuous.
\end{itemize}

The detailed rating criteria for each metric are shown in Figure~\ref{fig:rating_criteria}, and an example of the blind evaluation interface is provided in Figure~\ref{fig:eval_interface}.

\begin{figure}[t]
    \centering
    \includegraphics[width=\linewidth]{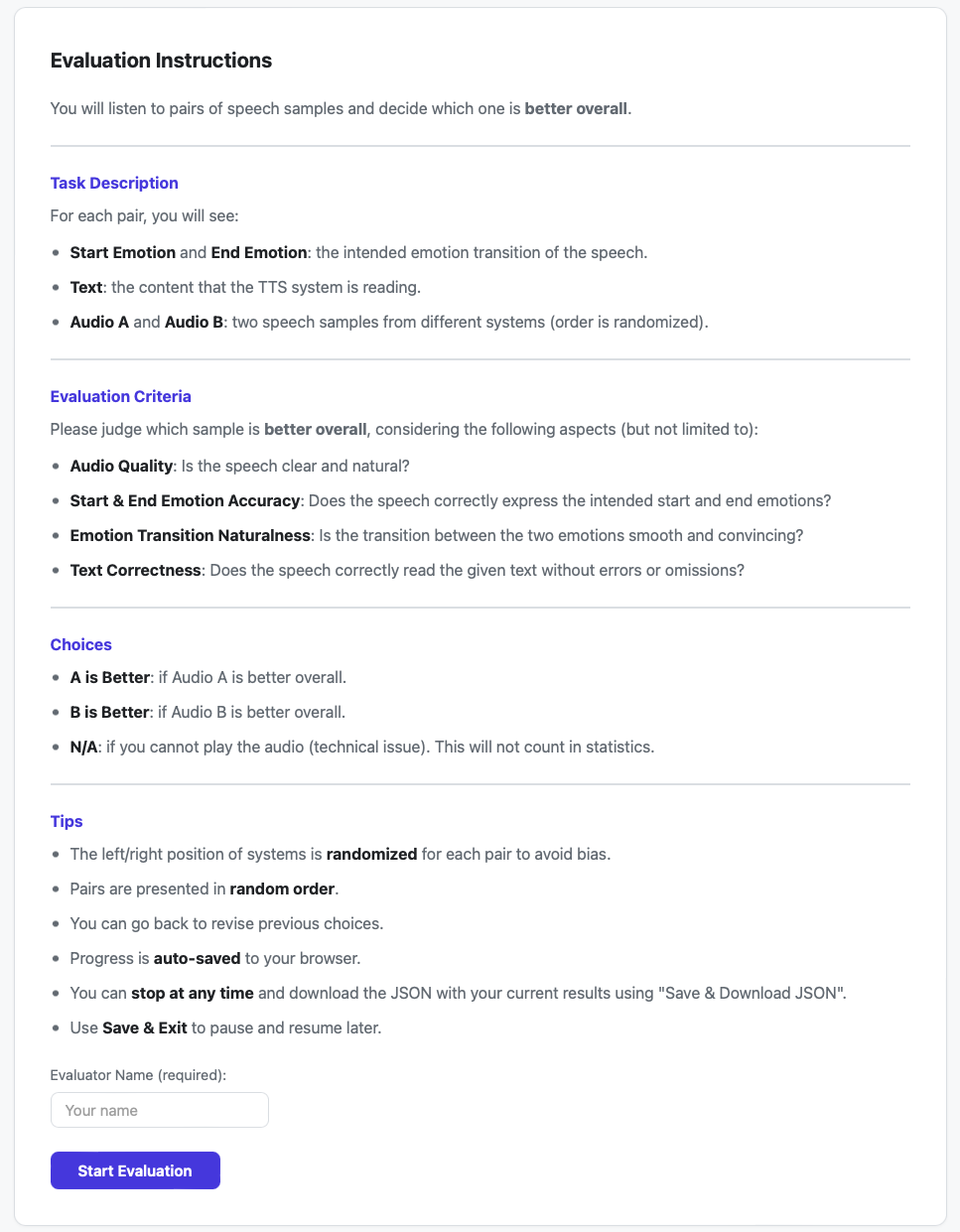}
    \vspace{-0.1 in}
    % \textcolor{red}{[TODO: Insert rating criteria screenshot --- 1--5 scale descriptions for MOS, MOS-Emo, MOS-Tra]}
    \caption{Rating criteria for the paired  evaluation.}
    \label{fig:rating_criteria2}
    \vspace{-0.1 in}
\end{figure}

\subsection{Paired Preference Evaluation Protocol}
\label{sec:paired_protocol}
For the paired preference evaluation in  \S\ref{sec:Pairwise}, we retain 4 raters from the original pool in  Appendix \ref{sec:MOSProtocol}  and include 7 additional raters, resulting in 11 raters for this study. For each system pair, 40 utterances were randomly sampled from the 200 utterances mentioned above, forming four paired comparison groups (160 pairs in total). Each rater completed about 85 comparisons on average, yielding about 6-fold repeated coverage per pair.

The detailed rating criteria for each metric are shown in Figure~\ref{fig:rating_criteria2}, and an example of the blind evaluation interface is provided in Figure~\ref{fig:eval_interface2}.

\subsection{Comparison with Commercial Systems}
\label{sec:commercial_protocol}

\begin{figure}[h]
    \centering
        \vspace{-0.1 in}
    \includegraphics[width=\linewidth]{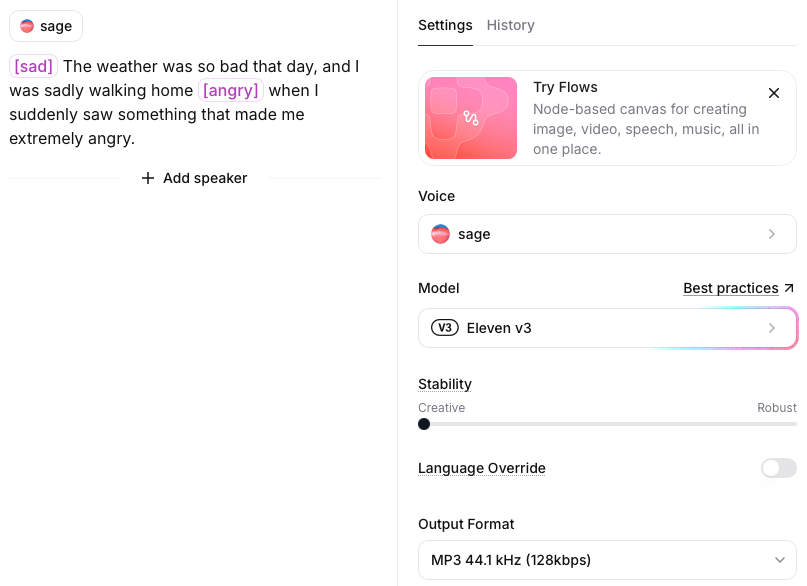}
    \caption{ElevenLabs~v3 inference interface used in our comparison, showing the manually inserted discrete audio tags (e.g., \texttt{[sad]}, \texttt{[angry]}) and the ``Stability'' control set to ``Creative''.}
    \label{fig:elevenlabs_interface}
    \vspace{-0.05 in}
\end{figure}

\begin{figure}[t]
    \centering
    \includegraphics[width=\linewidth]{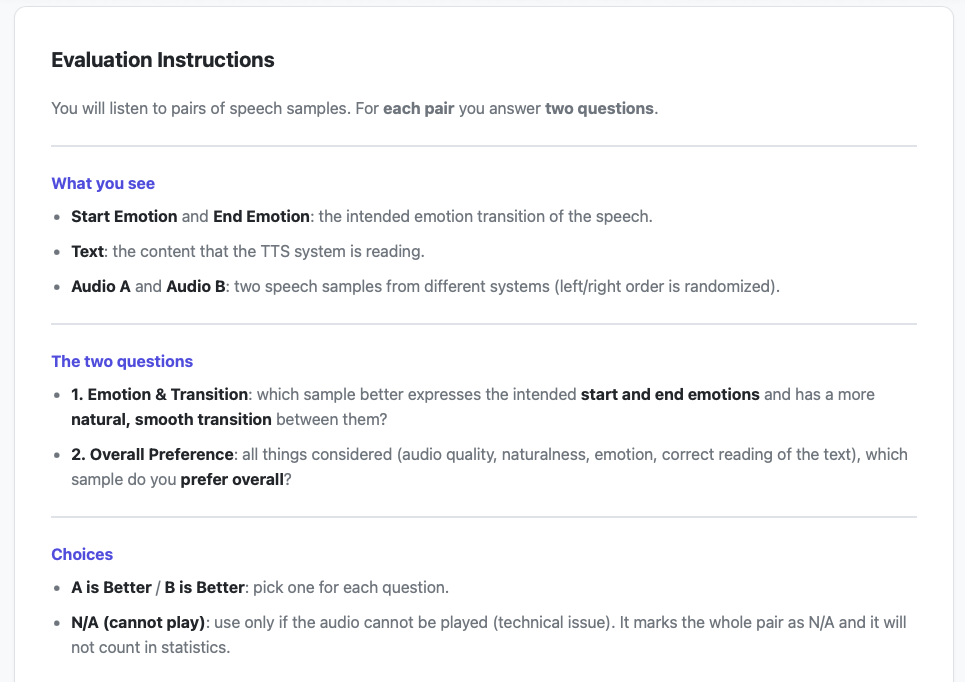}
    \vspace{-0.2 in}
    \caption{Rating criteria for the comparison with commercial systems.}
    \label{fig:rating_criteria3}
    % \vspace{-0.1 in}
\end{figure}

\begin{figure}[t]
    \centering
    \includegraphics[width=\linewidth]{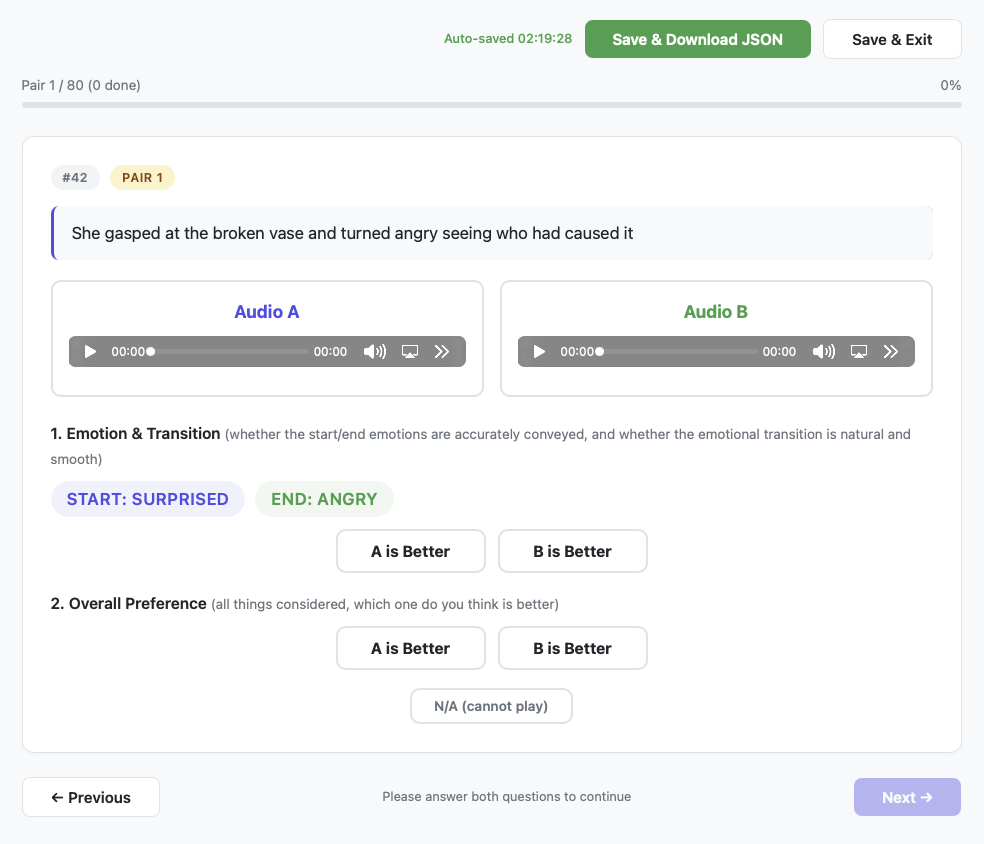}
    \vspace{-0.2 in}
    \caption{Screenshot of the paired preference evaluation interface. System identities are hidden, and the system order is randomly shuffled for each sample.}
    \label{fig:eval_interface3}
    \vspace{-0.1 in}
\end{figure}

This appendix details the two commercial systems compared in \S\ref{sec:commercial_main} and the evaluation protocol.
\textbf{GPT-4o mini TTS} (OpenAI's GPT-4o audio family) is controlled via a natural-language instruction prompt.
\textbf{ElevenLabs~v3} does not support natural-language prompts for emotion; instead, the user inserts discrete audio tags (e.g., \texttt{[happy]}, \texttt{[sad]}) at chosen positions\footnote{Prompting guide: \url{https://elevenlabs.io/docs/overview/capabilities/text-to-speech/best-practices\#prompting-eleven-v3-alpha}} \footnote{(updated Jul 5, 2026): \url{https://elevenlabs.io/blog/eleven-v3-situational-awareness}}, which requires an additional human or LLM step to place the tags and specifies discrete switches rather than continuous transitions. To present v3 in its best light, we placed the tags manually and set the ``Stability'' control to ``Creative'' (Figure~\ref{fig:elevenlabs_interface}), the most expressive setting per the official guidance.\footnote{\url{https://elevenlabs.io/app/speech-synthesis/text-to-speech}}

We ran a paired evaluation on 40 sentences per system (80 pairs in total) with 12 raters (4 newly recruited). Each rater evaluated about 53 pairs, and each pair was rated about 8 times. To avoid conflating synthesis quality with transition quality, we collected dimension-specific preferences: (i)~emotion accuracy and transition smoothness, and (ii)~overall preference as shown in Figure~\ref{fig:rating_criteria3} and Figure~\ref{fig:eval_interface3}.

\subsection{Inference Efficiency Evaluation Setup}
\label{sec:statis_infer}
This appendix summarizes the test set used for the inference efficiency evaluation. The evaluation set contains 100 sentences, with 5 additional warm-up sentences excluded from the statistics. The average sentence length is 13.9 words, with a standard deviation of 3.4 words. Sentence lengths range from 9 to 21 words, with a median of 13 words. In total, the test set contains 1{,}391 words.

\subsection{Test Set Generation Prompt}
\label{sec:test_prompt}

The following prompt was used with ChatGPT-5.2 to generate 200 test sentences with emotion transition annotations.

\begin{tcolorbox}[
  colback=gray!5,
  colframe=gray!60,
  fonttitle=\bfseries\small,
  title=Prompt for Test Set Generation,
  breakable,
  fontupper=\scriptsize\ttfamily,
  left=4pt, right=4pt, top=4pt, bottom=4pt
]
Task: Generate English sentences that describe a transition from one emotional state to another within a short narrative.\\[4pt]
Each sample must include: the sentence, the starting emotion, the ending emotion, and the VAD (Valence, Arousal, Dominance) values for both emotions.\\[6pt]
\textbf{EMOTION CATEGORIES}\\
angry, happy, sad, surprised, fearful, neutral\\
Transitions may occur between any pair of different emotions (30 possible pairs).\\[6pt]
\textbf{SENTENCE REQUIREMENTS}\\
Each sentence should describe a scenario or event that naturally causes an emotional shift. The narrative must clearly reflect the transition so that a speaker reading it aloud could express the emotional shift through prosody and tone.\\[2pt]
AVOID sentences that just name emotions. Instead, describe an event or realization that triggers the shift.\\
BAD: ``I was sad then I became happy''\\
GOOD: ``The long empty street made him feel alone until distant music drifted through the night''\\[2pt]
Sentence length: 12--30 words. Natural pauses allowed (commas, conjunctions). May include connectors such as: but, until, then, suddenly, moments later, after that, and then, when.\\[2pt]
Sentences should include diverse narrative contexts such as: daily life, workplace, social interaction, unexpected discoveries, environmental changes, memories, misunderstandings, surprising events, physical sensations, weather, animals, travel, etc.\\
Avoid repeating similar structures, scenarios, or phrasings.\\[6pt]
\textbf{VAD ANNOTATION}\\
Each emotion must include VAD values: (Valence, Arousal, Dominance). All values should be formatted to two decimal places.\\[2pt]
Target VAD Ranges:\\
\begin{tabular}{@{\hspace{2pt}}l@{\hspace{6pt}}l@{\hspace{6pt}}l@{\hspace{6pt}}l@{}}
angry:     & V (0.18--0.26) & A (0.82--0.90) & D (0.78--0.89) \\
happy:     & V (0.80--0.92) & A (0.76--0.88) & D (0.70--0.88) \\
sad:       & V (0.18--0.26) & A (0.18--0.28) & D (0.18--0.26) \\
surprised: & V (0.60--0.80) & A (0.80--0.89) & D (0.50--0.68) \\
fearful:   & V (0.18--0.26) & A (0.78--0.88) & D (0.18--0.28) \\
\end{tabular}\\[2pt]
Important: Values should reflect emotionally saturated states for non-neutral categories. Vary the values within the range; do not always use the same numbers.\\[2pt]
The VAD distance between starting and ending emotion must be clearly large enough to be perceptible. Avoid transitions where both emotions have very similar VAD values.\\[6pt]
\textbf{DISTRIBUTION CONSTRAINT}\\
Cover at least 25 unique emotion pairs out of 30. No single pair more than 15 times in 200 samples. Roughly balanced distribution.\\[6pt]
\textbf{FEW-SHOT EXAMPLES}\\
The loud crack startled him and then a creeping dread filled his chest | surprised | fearful | (0.68,0.86,0.60) | (0.20,0.84,0.22)\\
The cheerful laughter at the party was cut short when she saw the broken vase on the floor | happy | angry | (0.88,0.82,0.78) | (0.20,0.88,0.85)\\
A sudden gust of wind knocked the tent down and he scrambled in panic, but then burst into laughter | fearful | happy | (0.22,0.82,0.20) | (0.85,0.80,0.75)\\
She cursed the delayed train and then sat down feeling deeply miserable | angry | sad | (0.24,0.87,0.83) | (0.19,0.23,0.22)\\[6pt]
\textbf{OUTPUT FORMAT}\\
Each line: Sentence | emotion1 | emotion2 | (V,A,D) | (V,A,D)\\
Use | as separator. No numbering, no headers, no blank lines.\\[6pt]
\textbf{TARGET}: Generate 200 samples.
\end{tcolorbox}

\section{Emotion-Specific VAD Ranges}
\label{sec:vad_ranges}

\begin{table}[h]
\centering
\vspace{-0.2 in}
\caption{Emotion-specific VAD acceptance ranges for source corpus filtering. All values are on a $[0,1]$ scale. Ranges are intentionally broad to accommodate natural within-category variability.}
\label{tab:vad_ranges}
\resizebox{0.80\linewidth}{!}{
\begin{tabular}{@{}lccc@{}}
\toprule
\textbf{Emotion} & \textbf{Valence} & \textbf{Arousal} & \textbf{Dominance} \\
\midrule
Angry     & [0.00, 0.35] & [0.75, 1.00] & [0.70, 1.00] \\
Happy     & [0.65, 1.00] & [0.50, 1.00] & [0.55, 1.00] \\
Sad       & [0.00, 0.35] & [0.00, 0.50] & [0.00, 0.50] \\
Surprised & [0.35, 0.80] & [0.60, 1.00] & [0.35, 0.75] \\
Fearful   & [0.00, 0.45] & [0.40, 0.90] & [0.10, 0.55] \\
Disgusted & [0.00, 0.35] & [0.30, 0.70] & [0.30, 0.65] \\
Neutral   & [0.40, 0.60] & [0.40, 0.60] & [0.40, 0.60] \\
\bottomrule
\end{tabular}}
% \vspace{-0.1 in}
\end{table}

Table~\ref{tab:vad_ranges} lists the VAD acceptance ranges used for cross-validated filtering of the source corpus (\S\ref{sec:experiments}).
These ranges are \textbf{intentionally set broader} than prototypical VAD centroids~\cite{russell1980,mehrabian1996} for two reasons:
(1)~natural emotional speech exhibits substantial within-category variability~\cite{scherer2003,banse1996}, and
(2)~the VAD predictor~\cite{wagner2023dawn} has estimation uncertainty, so tight boundaries would conflate predictor error with genuine mismatch.
The filtering retains samples only if their categorical label \emph{and} predicted VAD are mutually consistent, removing mislabeled or acoustically ambiguous utterances.

\section{Objective Smoothness Evaluation}
\label{sec:objective}

\begin{table}[h]
% \vspace{-0.1 in}
\centering
\caption{Prosodic Jerk Ratio for the proposed EmoTra-TTS and baselines in Categories (d) and (e) of Table~\ref{tab:comparison}. All differences between EmoTra-TTS and the baselines are statistically significant ($p < 0.001$).}
\vspace{-0.05 in}
\small
\resizebox{0.8\linewidth}{!}{
\begin{tabular}{lc}
\toprule
\textbf{System} & \textbf{JR-F0} $\downarrow$ \\
\midrule
\textbf{EmoTra-TTS (Ours) }    & \textbf{0.025 $\pm$ 0.009} \\
Synthetic data (Ours) & 0.030 $\pm$ 0.014 \\
Qwen3-TTS~\cite{Qwen3TTS}             & 0.036 $\pm$ 0.016 \\
WeSCon~\cite{wescon}                & 0.038 $\pm$ 0.017 \\
MOSS-TTS~\cite{mosstts}            & 0.038 $\pm$ 0.017 \\
CosyVoice2~\cite{cosyvoice2}         & 0.039 $\pm$ 0.018 \\
\bottomrule
\end{tabular}
}

\label{tab:jr_f0}
\vspace{-0.10 in}
\end{table}

To complement the perceptual evaluation with an objective measure, we introduce the \emph{Prosodic Jerk Ratio} (JR-F0), a signal-level metric computed directly from $f_0$ via Praat. Given a voiced $f_0$ contour, we measure the second-order difference $\Delta^{2} f_0[n] = f_0[n{+}1] - 2f_0[n] + f_0[n{-}1]$ and define

\begin{equation}
\mathrm{JR\text{-}F0} \;=\;
\frac{\#\{\,n : |\Delta^{2} f_0[n]| > \tau\,\}}
     {\#\{\,n : f_0[n]~\text{voiced}\,\}},
\end{equation}
i.e.\ the fraction of voiced frames with abrupt pitch discontinuities, where $\tau$ is an adaptive per-utterance threshold. Lower JR-F0 indicates smoother prosodic evolution.

\begin{table*}[t]
\centering
\caption{Full injection architecture comparison across all seven variants. SIM: speaker similarity. WER: word error rate (\%). We decompose MOS into MOS-Qua, MOS-Emo, and MOS-Tra for a more fine-grained and independent evaluation of speech naturalness, emotion rendering adequacy, and intra-utterance emotion transition smoothness, respectively. $^\dag$Finetune is excluded from blind test evaluation due to severely degraded content (decoder drift renders most outputs unintelligible), precluding meaningful perceptual assessment.}
% \vspace{-0.1 in}
\label{tab:ablation_full}
\resizebox{\textwidth}{!}{
\begin{tabular}{@{}llccclrccl@{}}
\toprule
\multirow{2}{*}{\textbf{Variant}} & \multirow{2}{*}{\textbf{Injection}} & \multirow{2}{*}{\textbf{$\Delta$Params}} & \multirow{2}{*}{\textbf{Decoder}} & \multicolumn{3}{c}{\textbf{Quality Evaluation}} & \multicolumn{2}{c}{\textbf{Emotion Evaluation}} & \multirow{2}{*}{\textbf{Key finding}} \\
\cmidrule(r){5-7} \cmidrule(r){8-9}
 & & & & MOS-Qua$\uparrow$ & SIM$\uparrow$ & WER$\downarrow$ & MOS-Emo$\uparrow$ & MOS-Tra$\uparrow$ & \\
\midrule
Finetune$^\dag$ & Linear & 72K & Trainable & --- & --- & --- & --- & --- & Decoder drift \\
\textbf{Linear} & \textbf{Linear, additive} & 82K & Frozen & \textbf{3.60}$\pm$0.12 &\textbf{0.707} & \textbf{1.65} &  3.39$\pm$0.15 & 3.33$\pm$0.16 &Emotion Capacity limit \\
MLP+LoRA & MLP+LoRA & 2.9M & LoRA       & 1.03$\pm$0.05 & 0.627 & 107.15 & 1.69$\pm$0.13 & 1.42$\pm$0.11 & Content collapse \\
MLP-Add & MLP, additive & 280K & Frozen & 2.10$\pm$0.17 & 0.595 & 20.09 & 2.49$\pm$0.17 & 2.17$\pm$0.17 & Magnitude mismatch \\
FiLM & FiLM modulation & 560K & Frozen & 1.96$\pm$0.16 & 0.400 & 53.35 & 2.44$\pm$0.17 & 1.82$\pm$0.16 & MLP compensation \\
Norm Clamp & Norm Clamp & 280K & Frozen & 3.29$\pm$0.15  & 0.692 & 3.89 & 3.30$\pm$0.15 & 3.02$\pm$0.16 &  Gradient suppression \\
\textbf{Dir--Mag (ours)} & \textbf{LN+fixed} $\epsilon$ & {280K} & {Frozen} & 3.54$\pm$0.13 &  0.701 & 1.77 &  \textbf{3.52}$\pm$0.15 & \textbf{3.63}$\pm$0.17 & \textbf{Dir--mag decoupled} \\
\bottomrule
\end{tabular}
}
% \vspace{-0.1 in}
\end{table*}

As shown in Table~\ref{tab:jr_f0}, EmoTra-TTS attains the lowest JR-F0, corroborating the perceptual transition-naturalness results with a fully signal-level measure. Notably, EmoTra-TTS is even smoother than its own synthetic training data ($0.025$ vs.\ $0.030$): the synthetic corpus retains residual discontinuities from four-segment blending (\S\ref{sec:multipass}), whereas frame-level interpolated VAD conditioning yields globally continuous prosody.

\section{Detailed Failure Analysis}
\label{sec:failure_analysis}

This appendix documents the iterative development across all seven injection architectures (Table~\ref{tab:ablation_full}), each failure revealing a distinct phenomenon in auxiliary conditioning of frozen generative models.

\subsection{Decoder Drift (Finetune)}
A linear projection ($896 \!\rightarrow\! 80$; $\sim$72K parameters) injects emotion through the conditioning channel $\bm{c}$ with the decoder \emph{jointly trained}.
Strong emotion is achieved, but the decoder's $\sim$639M parameters absorb the adaptation by shifting pretrained weights, degrading content fidelity.
This establishes the \emph{decoder freezing} principle for all subsequent variants.

\subsection{Emotion Capacity Limit (Linear)}
A single linear layer ($1024 \!\rightarrow\! 80$; 82K parameters) with frozen decoder and additive injection preserves content (MOS-Qua 3.60) but produces only moderate emotion (MOS-Emo 3.39, MOS-Tra 3.33).
The linear mapping lacks capacity for the nonlinear $d_\text{SER}$-to-$d_\text{mel}$ transformation~\cite{alain2017understanding}, establishing a \emph{lower bound} on model complexity for expressive conditioning.

\subsection{Content Collapse (MLP+LoRA)}
A two-layer MLP ($1024 \!\rightarrow\! 256 \!\rightarrow\! 80$; $\sim$280K) with LoRA~\cite{lora} rank-8 adapters on all 336 decoder layers ($\sim$2.9M total) yields emotion but \emph{zero} intelligible content (MOS-Qua 1.03, WER 107.15\%, SIM 0.627).
LoRA's global perturbation shifts the flow decoder's content--style equilibrium, catastrophically amplifying style at the expense of content, paralleling findings that adapter placement critically affects multi-conditional generation~\cite{hu2024animate,ctrladapter}.
This establishes that {decoder weights should not be modified for single-axis adaptation}, even with parameter-efficient methods~\cite{lora}.

\subsection{Magnitude Mismatch (MLP-Add)}
\label{sec:magnitude_results}
The same MLP without LoRA ($\sim$280K parameters, decoder frozen) achieves strong emotion but content failure (MOS-Qua 2.10, WER 20.09\%) with a diagnostic {bimodal} pattern: emotion is reflected in both successful and garbled samples, indicating a distributional out-of-range problem.
Norm measurements reveal the cause: $\|\bm{e}_\text{emo}(t)\|_2 \approx 1.248 \approx 2\times \|\bm{e}\|_2$, pushing the combined embedding $3\times$ beyond the decoder's pretrained operating range.

We isolate magnitude as the causal factor via inference-time manual scaling $\lambda$ on frozen MLP weights.
Content failure is monotonically related to the emotion-to-speaker norm ratio: at $\lambda\!=\!0.5$ (ratio $\approx 1.0$), content is fully preserved; at $\lambda\!=\!1.0$ (ratio $\approx 2.0$), only a small portion of samples remain intelligible.
Emotion direction is correct even in garbled samples, confirming that the problem is {exclusively} one of magnitude.
This directly validates our Dir--Mag design and generalizes: for {any} additive conditioning of frozen models, the conditioning-to-existing-signal norm ratio must stay within the pretrained operating range.

\subsection{MLP Compensation (FiLM)}
FiLM~\cite{film} modulation ($\hat{\bm{e}}(t) = \bm{\gamma}(t) \odot \bm{e} + \bm{\beta}(t)$; $\sim$560K parameters, identity-initialized~\cite{controlnet}, decoder frozen) is designed to provide learnable per-dimension scale and shift.
However, FiLM fails even more severely than MLP-Add (MOS-Qua 1.96, WER 53.35\%, SIM 0.400): the MLP compensates by inflating output magnitude $3$--$5\times$ during joint optimization.
This demonstrates that \emph{soft constraints on entangled factors are insufficient}~\cite{locatello2019challenging}. The flow loss, averaging over all frames, provides no gradient penalizing magnitude growth, and the MLP and FiLM parameters co-adapt to circumvent the constraint.

\subsection{Gradient Suppression (Norm Clamp)}
Hard norm clamping ($\tau\!=\!0.70$, derived from the manual scaling experiment; $\sim$280K parameters, decoder frozen) substantially improves content over MLP-Add (MOS-Qua 3.29, WER 3.89\%), confirming magnitude control as the key factor.
However, emotion is only moderate (MOS-Emo 3.30) due to \emph{gradient suppression}: beyond $\tau$, clamping zeros the radial gradient component, and the discontinuous transition between clamped and unclamped regimes causes the MLP to oscillate near the boundary rather than converging smoothly.
This demonstrates that even a \emph{correct} threshold is insufficient without a \emph{smooth} normalization mechanism.

\subsection{Success: Direction--Magnitude Decoupled}
The same MLP followed by LayerNorm($d_\text{mel}\!=\!80$) and fixed $\epsilon\!=\!0.07$ ($\sim$280K + 160 parameters, decoder frozen) achieves content fidelity with strong emotion (MOS-Qua 3.54, MOS-Emo 3.52, MOS-Tra 3.63).
LayerNorm locks magnitude at $\sqrt{d_\text{mel}} \approx 8.94$; the effective norm $\epsilon \cdot \sqrt{d_\text{mel}} \approx 0.626$ matches the speaker norm ($\|\bm{e}\|_2 \approx 0.622$).
The MLP learns emotion \emph{directions} freely but cannot inflate \emph{magnitude}, confirming that \emph{structural} constraints are necessary for safe auxiliary conditioning.

\section{Crossfade Curve Selection}
\label{sec:crossfade_analysis}

\begin{table}[h]
\centering
\caption{Crossfade curves evaluated for mel-level blending. $\sigma_\kappa(t) = 1/(1+e^{-\kappa(t-0.5)})$, normalized to $[0,1]$; $s(t) = \frac{1}{2}(1-\cos\pi t)$. Act. fr. indicates active frames.}
\label{tab:crossfade_curves}
% \vspace{-0.05 in}
\resizebox{\columnwidth}{!}{
\begin{tabular}{@{}lllcc@{}}
\toprule
\textbf{Curve} & $w_A(t)$ & $w_B(t)$ & \textbf{Act. fr.} & $w_A^2{+}w_B^2{=}1$? \\
\midrule
Linear & $1 - t$ & $t$ & 200 & \ding{55} \\
Equal-power cosine & $\cos\!\big(\tfrac{\pi}{2}\,s(t)\big)$ & $\sin\!\big(\tfrac{\pi}{2}\,s(t)\big)$ & ${\sim}$150 & \ding{51} \\
{Sigmoid symmetric} & $1-\sigma_{12}(t)$ & $\sigma_{12}(t)$ & ${\sim}$34 & \ding{55} \\
Sigmoid asymmetric & $\sqrt{1-\sigma_{12}(t)}$ & $\sigma_{12}(t)^2$ & ${\sim}$34 & \ding{55} \\
Ultra-steep sigmoid & $\sqrt{1-\sigma_{100}(t)}$ & $\sigma_{100}(t)^{1.5}$ & ${\sim}$4 & \ding{55} \\
\bottomrule
\end{tabular}
}
\vspace{-0.05 in}
\end{table}

This appendix supplements the mel-spectrogram-level crossfade in \S\ref{sec:multipass} (Fig.~\ref{fig:systhe}) by documenting the curve selection process.
The crossfade curve is \textbf{not} a core contribution; we include this analysis for interested readers.
The selection was based on informal listening by the authors (${\sim}$20 samples per curve, not a formal blind test).

Within the transition region, blending follows $\text{mel}(t) = w_A(t)\,\text{mel}_A(t) + w_B(t)\,\text{mel}_B(t)$ on normalized $t\!\in\![0,1]$.
Since the quality of synthetic training data directly depends on the blending curve, we evaluated five candidates (Table~\ref{tab:crossfade_curves}) to find the best trade-off between transition smoothness and emotion purity for data generation.

\textbf{Linear} blending made the splice point audible in a subset of samples.
\textbf{Equal-power cosine} removed the splice artifact but noticeably degraded quality in the transition region.
\textbf{Ultra-steep sigmoid} ($\kappa{=}100$) compressed blending to ${\sim}$4 frames, yielding no artifacts but sounding perceptually identical to a hard switch and losing all sense of gradual transition.
Both \textbf{sigmoid} variants ($\kappa{=}12$) concentrate blending in ${\sim}$34 central frames while keeping edges binary and sound largely identical in practice.
The asymmetric variant exhibited mild spectral artifacts in only a few samples; the symmetric version avoided these, so we adopted it for consistency.

We adopted the \textbf{sigmoid symmetric} with $\kappa{=}12$ (Eq.~\eqref{eq:sigmoid}).
The key finding is that $\kappa$ governs the smoothness--purity trade-off: too gradual (linear, cosine) sacrifices emotion contrast; too steep ($\kappa{\geq}100$) collapses to a hard switch.

\end{document}